\documentclass[10pt,final,journal,twocolumn]{IEEEtran}

\usepackage{cite}
\usepackage{float}
\usepackage{colortbl,booktabs}
\usepackage{environ}
\usepackage{color,soul}
\usepackage{enumitem}
\usepackage{amsmath}
\usepackage{subfig}
\usepackage{graphicx}
\graphicspath{{figs/}}
\usepackage{float}
\usepackage{amsmath, amsthm, amssymb, amsfonts, mathtools}
\usepackage{bm}
\usepackage{textcomp}
\usepackage{xcolor}
\usepackage{float}
\usepackage{color}
\usepackage{todonotes}
\usepackage{diagbox}
\usepackage{amsmath}
\usepackage{tablefootnote}
\usepackage{xr}
\usepackage{textcomp}
\usepackage{xcolor}
\usepackage[hyphens]{url}
\usepackage{fancyhdr}
\usepackage{hyperref}

\newcommand\numberthis{\addtocounter{equation}{1}\tag{\theequation}}

\begin{document}

\title{BF-IMNA: A Bit Fluid In-Memory Neural Architecture for Neural Network Acceleration}
\author{Mariam Rakka, Rachid Karami, Ahmed M. Eltawil, Mohammed E. Fouda, and Fadi Kurdahi
\thanks{M. Rakka, R. Karami and F. Kurdahi are with University of California-Irvine, Irvine, CA, USA 92697-2625. Email: mrakka@uci.edu, A. Eltawil Is with King Abdullah University of Science and Technology (KAUST), Thuwal 23955, Saudi Arabia. M. Fouda is Rain Neuromorphics Inc., San Fransisco, CA, USA. Email:foudam@uci.edu}
}


\maketitle

\pdfpagewidth=8.5in
\pdfpageheight=11in





\begin{abstract}
Mixed-precision quantization works defining per-layer, per-network, per-channel, or per-parameter precisions for Neural Networks (NNs) are gaining traction for their efficient realization on the hardware leading to higher throughput and lower energy. In-Memory Computing (IMC) accelerator architectures like PUMA \cite{ankit2019puma}, PipeLayer \cite{song2017pipelayer}, and ISAAC \cite{shafiee2016isaac}, are offered as alternatives to traditional architectures relying on a data-centric computational paradigm, diminishing the memory wall problem, and scoring high throughput and energy efficiency. These accelerators can support static fixed-precision but are not flexible to support mixed-precision NNs.

In this paper, we present BF-IMNA, a bit fluid IMC accelerator for end-to-end Convolutional NN (CNN) inference that is capable of static and dynamic mixed-precision without any hardware reconfiguration overhead at run-time. At the heart of BF-IMNA are Associative Processors (APs), which are bit-serial word-parallel Single Instruction, Multiple Data (SIMD)-like engines. We report the performance of end-to-end inference of ImageNet on AlexNet, VGG16, and ResNet50 on BF-IMNA for different technologies (eNVM and NVM), mixed-precision configurations, and supply voltages. To demonstrate bit fluidity, we implement HAWQ-V3's \cite{yao2021hawq} per-layer mixed-precision configurations for ResNet18 on BF-IMNA using different latency budgets, and results reveal a trade-off between accuracy and Energy-Delay Product (EDP): On one hand, mixed-precision with a high latency constraint achieves the closest accuracy to fixed-precision INT8 and reports a high (worse) EDP compared to fixed-precision INT4. On the other hand, with a low latency constraint, BF-IMNA reports the closest EDP to fixed-precision INT4, with a higher degradation in accuracy compared to fixed-precision INT8. 
We also show that BF-IMNA with fixed-precision configuration still delivers performance that is comparable to current state-of-the-art accelerators: BF-IMNA achieves  $20\%$ higher energy efficiency compared to PipeLayer and $2\%$ higher throughput compared to ISAAC.
\end{abstract}


\section{Introduction}
Deep Learning (DL) algorithms, especially Convolutional Neural Networks (CNNs), have gained widespread popularity, playing a pivotal role in applications like computer vision, natural language processing, and recommendation systems \cite{liu2017survey,goel2020survey, dou2020exploiting, sinha2022dnn}. Deploying and scaling CNNs on resource-constrained edge devices amid Moore's law stagnation and the rise of IoT applications requires maximizing performance while minimizing computational overhead \cite{li2018edge, theis2017end, hussain2022design}. To achieve this "Pareto" optimality, literature explores a mix-and-match plethora of techniques including hardware customization and algorithmic optimizations.

Hardware customization encompasses developing accelerator platforms that maximize efficiency and throughput. Examples of State-of-The-Art (SOTA) specialized von Neumann architectures include the Tensor Processing Unit (TPU) which relies on systolic array architectures \cite{tpuv4}, Eyeriss which relies on two-dimensional spatial arrays of processing elements \cite{chen2019eyeriss}, and the architecture in \cite{sim201614} which has dedicated dual-range Multiply-Accumulate (MAC) blocks. Beyond von Neumann, PUMA \cite{ankit2019puma}, ISAAC \cite{shafiee2016isaac}, PipeLayer \cite{song2017pipelayer}, and \cite{valavi201964, yantir2021imca} leverage In-Memory Computing (IMC)-based paradigms for CNN acceleration, relying on analog/digital computing. IMC is a data-centric computational paradigm that abolishes the memory wall problem suffered by Von-Neumann by allowing computation to be performed inside memory \cite{ma2020memory}.

Mixed-precision quantization has recently emerged as an algorithmic optimization that enables a fast, low-memory footprint deployment of CNNs without compromising performance. With mixed-precision quantization, the model's weights/activations are quantized to a certain precision chosen from a set of possible bitwidths, in a way that the overall mixed-precision configuration yields high accuracy. The granularity of mixed-precision can be fine-grained (at the weight/activation level; each weight/activation can be assigned a precision) or coarse-grained  (at the layer level; all weights/activations in one layer are assigned the same precision, but precisions vary between layers). Mixed-precision CNN frameworks\cite{rakka2022mixed} are categorized based on their optimization techniques: 1-) gradient-based works like Bayesian Bits \cite{van2020bayesian}, 2-) Reinforcement Learning-based works like AutoQ \cite{lou2019autoq}, 3-) heuristic-based works like HAWQ-V3 \cite{yao2021hawq}, and 4-) meta-heuristic-based works like APQ \cite{wang2020apq}. Several works concluded by calling for hardware-aware mixed-precision frameworks; capable of accommodating run-time changing requirements of the target platforms and call for "hardware platforms capable of supporting dynamic mixed-precision, where the precision can be changed at run-time" \cite{rakka2022mixed}.

All of the aforementioned SOTA CNN accelerators were designed with fixed-precision (all weights/activations have the same precision across all models). Even if the same design rules/considerations of these accelerators were followed to accommodate for mixed-precision, those accelerators will end up supporting one mixed-precision configuration which is chosen statically at design time. This is not practical as different mixed-precision works have shown that the optimal configuration changes with the model \cite{yao2021hawq,lou2019autoq, wang2019haq}. Moreover, such accelerators would not be able to accommodate changing requirements at run-time which would require a change in the mixed-precision configuration \cite{bulat2021bit}. 

While the list of hardware accelerators that support fixed-precision CNN inference is long, the literature currently lacks end-to-end hardware accelerators tailored for mixed-precision CNN inference, especially dynamic mixed-precision. SOTA mixed-precision CNN accelerators include Stripes \cite{judd2016stripes} and BitFusion \cite{sharma2018bit}. While Stripes is a bit-serial CNN accelerator that enables mixed-precision, it is not an end-to-end accelerator design but rather focuses on accelerating only convolution. BitFusion is a more recent, spatial, end-to-end CNN accelerator capable of bit-level fusion and outperforms Stripes \cite{sharma2018bit}.

\begin{figure}[t!]
{\includegraphics[width=1\columnwidth]{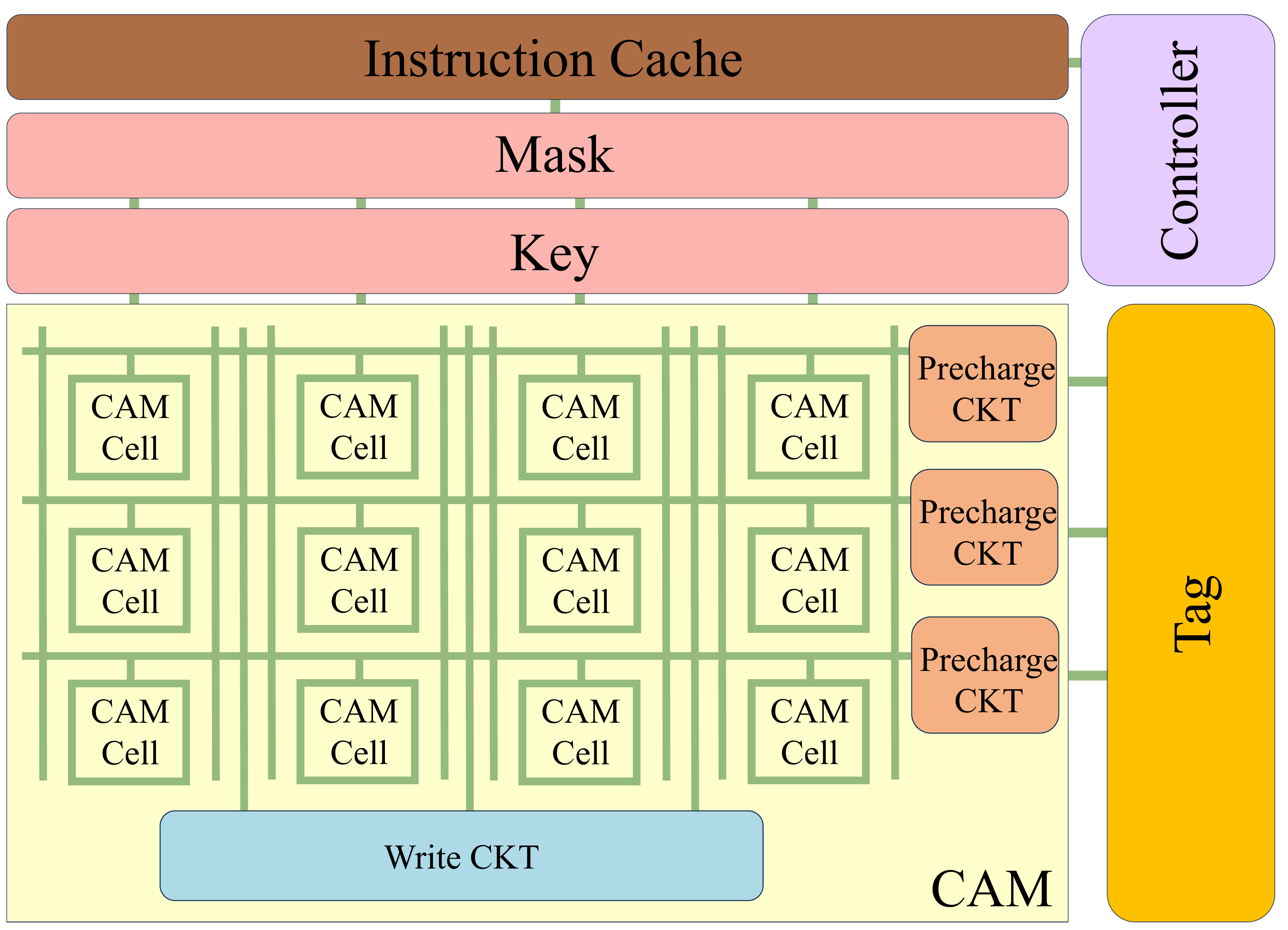}}
\centering
\caption{General architecture of an AP.\label{xbarap}}
    \vspace{-0.15in}
\end{figure}

In this paper, we propose BF-IMNA: A \underline{B}it \underline{F}luid
\underline{I}n-\underline{M}emory \underline{N}eural \underline{A}rchitecture for end-to-end full-fledged fixed and dynamic CNN acceleration. Bit fluidity is a term we henceforth use to describe the ability to enable mixed-precision inference. Unlike BitFusion, BF-IMNA relies on the \textit{IMC-based} Associative Processor (AP) as its building block. The AP shown in Fig. \ref{xbarap} relies on a Content Addressable Memory (CAM) to carry out massively parallel logical/arithmetic operations in a bit-serial word parallel manner \cite{fouda2022memory, krikelis1994associative}. The bit-serial operation inherently supports mixed-precision without the need of dynamic reconfigurability. CAMs are IMC architectures utilized in several applications that require parallel searching like IP routing, text analytics, data mining, and data analytics \cite{karam2015emerging,zha2020hyper, foster1976content}. APs perform arithmetic and logical operations by a series of search/write stages following the operation's Look-Up Table (LUT) applied in a bit-serial word-parallel manner (i.e. between pairs of columns of the CAM). An extension of the AP is a 2D AP proposed in \cite{yantir2018two} which provides the capability of performing logical/arithmetic operations between pairs of rows in addition to being able to perform those between pairs of columns. The 1D and 2D APs are the only building blocks of BF-IMNA: we use 1D APs as Memory APs (MAPs); the on-chip memory storage, and we use 2D APs as Computation APs (CAPs), the on-chip computational units capable of performing any CNN operation. The architecture of BF-IMNA is shown in Fig. \ref{arch}. Our contributions are below.

\noindent\textbf{1-} We formulate the complexities of performing CNN-related operations such addition, multiplication, ReLU activation, average pooling, and max pooling on the 2D AP, taking into consideration the data movements. 

\noindent\textbf{2-} We map the end-to-end inference operations into AP blocks, and we present, based on different hardware configurations, two BF-IMNA designs: one that offers maximum parallelism and another with limited resources.

\noindent\textbf{3-} We develop an in-house simulator for BF-IMNA that estimates the performance metrics (energy, area, latency) of end-to-end inference of ImageNet on AlexNet, ResNet50, and VGG16. Then we carry out a design space exploration for different CAM cell technologies (SRAM/ReRAM), different per-layer mixed-precision configurations, and different supply voltages.

\noindent\textbf{4-} We demonstrate BF-IMNA's bit fluidity by implementing HAWQ-V3's \cite{yao2021hawq} per-layer mixed-precision configurations for different 
latency budgets. The different mixed-precisions reveal a trade-off between accuracy and Energy-Delay Product (EDP). 

\noindent\textbf{5-} We compare BF-IMNA using fixed-precision configurations to SOTA accelerators. Our results demonstrate that BF-IMNA achieves comparable results even at a high fixed-precision (16 bits) where APs start to lose their advantage: $1.02\times$ higher
throughput with $3.66\times$ lower energy efficiency and $1.19\times$ higher energy efficiency with $2.95\times$ lower throughput compared to ISAAC and PipeLayer.

The rest of the paper is organized as follows: Section II provides background, and section III elaborates on BF-IMNA. Sections IV and V present the implementation details, results, and discussion. Section VI concludes the paper.

\section{Background}
\subsection{Convolutional Neural Networks}
CNNs, a subset of DL widely used in computer vision \cite{li2021survey}, typically incorporate convolutional, pooling, and fully connected layers. Convolutional layers perform high-dimensional convolutions on input feature maps using weighted filters. After convolution, an activation function, such as ReLU, introduces non-linearity to generate the layer's output feature map. Pooling layers (max and average) employ sliding filters with a $z \times z$ size and a stride of $S_t$. Max pooling selects the maximum data point in the window, while average pooling calculates the mean. Fully connected layers involve linear matrix-vector-multiplication connecting all inputs to all outputs.

\subsection{Two Dimensional Associative Processors}
The hardware architecture of a 2D Computational AP (CAP) is depicted in Fig. \ref{arch}, comprising a 2D CAM for data storage whose building block cells are either SRAM-based or ReRAM-based. Note that compared to the cells of 1D APs (MAPs), the cell designs in 2D APs incorporate an additional nmos/pmos device for each ReRAM/SRAM, controlled by vertical search lines (S(V)s) to enable vertical search.

\noindent\textbf{Arithmetic/Logical Operations:} An arithmetic/logical operation is performed in the 2D AP by following a sequence of compare-write stages applied to the CAM's data, following the operation's LUT. Particularly, the controller receives logical/arithmetic operations to be performed on the data stored in the 2D CAM and translates those into key-mask values, which will be used in the compare-write stages following the operation's LUT. The key register stores the value that is written or compared against, and the mask register indicates bit(s) that is(are) activated during comparison or write. The compare stage is a search applied to the CAM's data. To facilitate search, each row/column terminates in precharge circuitry, and a capacitor that feeds into a sense amplifier connected to the tag. Tag registers indicate matched rows/columns during a search and facilitate bit-sequential/word-sequential reading. The 2D AP supports vertical and horizontal search, where search drivers drive cells of interest.

\noindent\textbf{Reading Modes:} Two reading modes are supported in the 2D AP: bit-sequential and word-sequential. For bit-sequential reading, all columns are masked except the one of interest, then the key corresponding to that column is set to "one", which is then used by the corresponding search driver to drive the search lines of that column. Since cells in the column of interest match with the input "one" if they store one (hence setting the corresponding tag registers) or mismatch with it if they store a zero (hence resetting the corresponding tag registers), the tag registers will eventually hold the values that are stored in that column. The read column (now stored in the tag) can be used to rewrite another column in the CAM (via the write driver) or to be passed to a general-purpose register with the help of the interconnection interface. Word-sequential reading is done via a search operation similar to bit-sequential mode, but relying on the vertical search lines, vertical key and mask registers, and the horizontal tag registers.

\noindent\textbf{Writing Modes:} Writing occurs in bit-sequential or word-sequential mode, with a two-cycle requirement per writing a row/column. The 2D-AP presented in \cite{yantir2018efficient} enables vertical segmentation, treating each group of rows as a separate 2D AP. Segmentation reduces operation complexity and offers energy efficiency but limits hardware programmability. To render our design more general-purpose, instead of using vertical segmentation, we assume that we can perform operations on multiple 2D APs that operate in parallel.

\begin{figure}[hbt!]
    \centering
{\includegraphics[width=0.9\columnwidth]{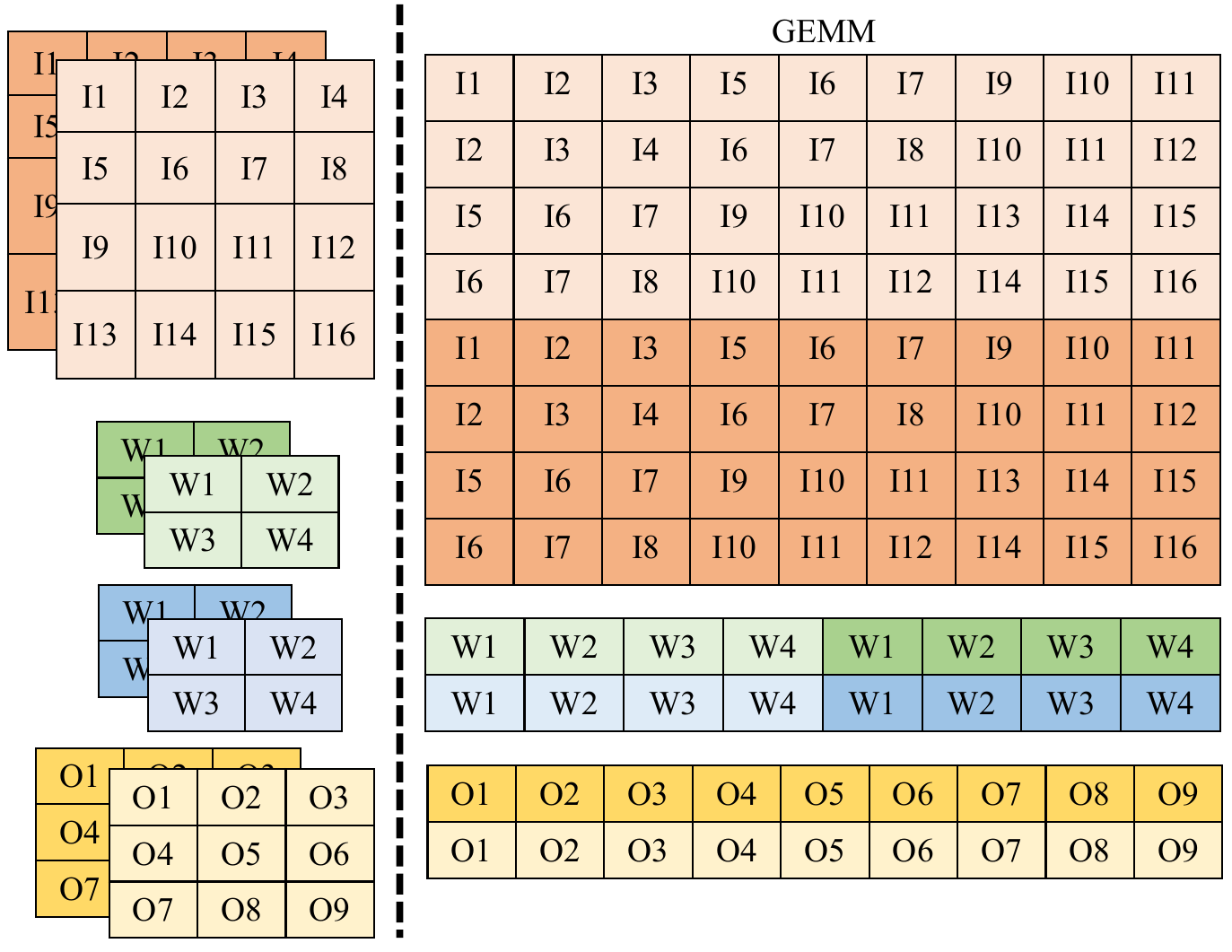}}
\caption{Convolutional layer example with a $2\times 2 \times 2$ input and a $2 \times 2 \times 2 \times 2$ filter and its corresponding Input P, Kernel K, and output matrices after GEMM transformation.}
\label{convex}
\vspace{-0.15in}
\end{figure}

\begin{figure}[ht!]
\centering   {\includegraphics[width=1\columnwidth]{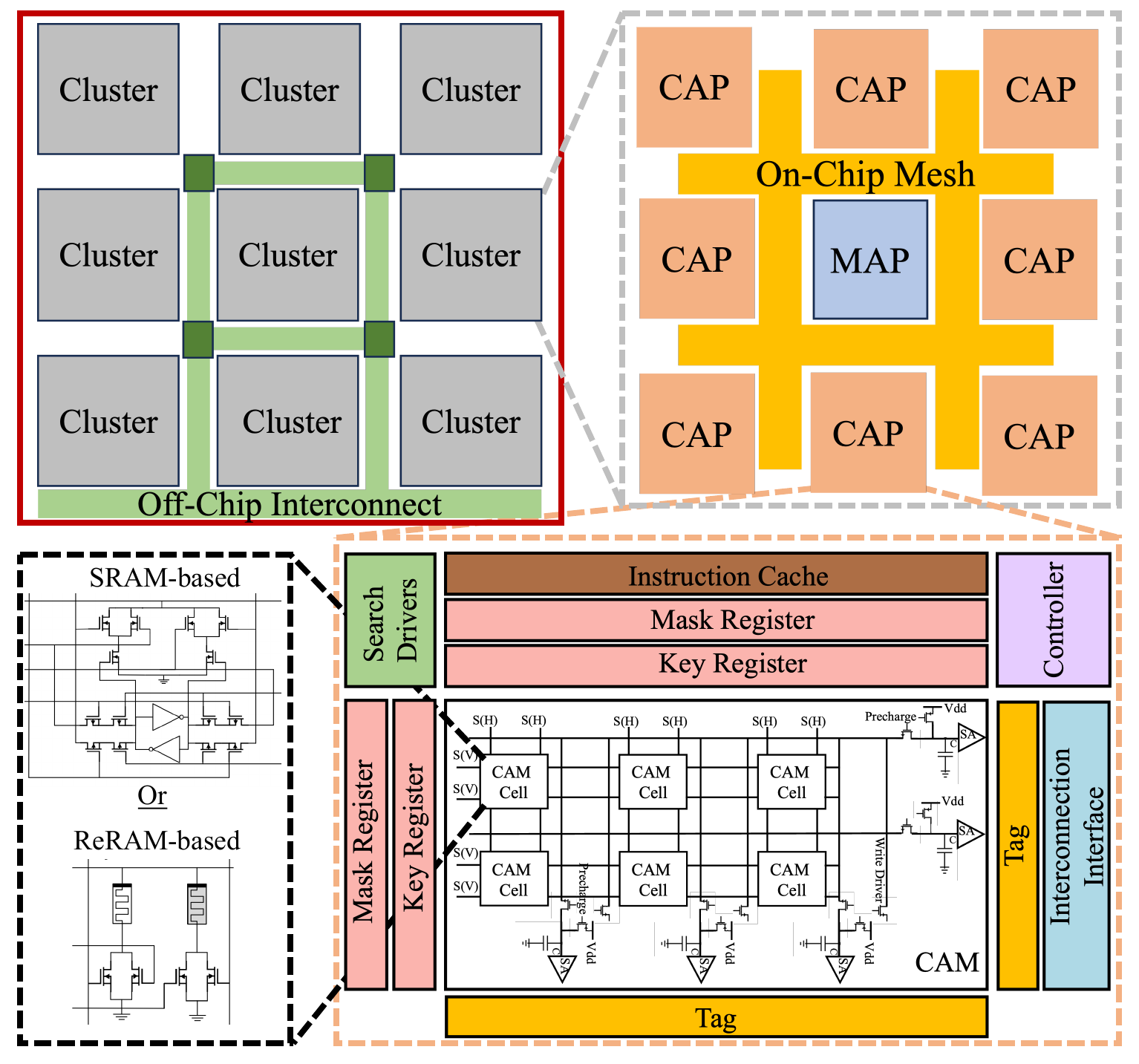}}
    \caption{BF-IMNA architecture.}
    \label{arch}
\end{figure}
\subsection{General Matrix Multiply}
In our proposed architecture, a convolution is mapped onto the APs using General Matrix Multiply (GEMM), which implements the convolution by transforming one input matrix into a Toeplitz matrix. This transformation involves replicating image pixels across matrix columns, and the convolution is executed through a general matrix multiplication approach known as "im2col", widely utilized in DL frameworks like Caffe and Torch \cite{vasudevan2017parallel, jia2014caffe, collobert2002torch}. Given an input size of $\{height, width, channels\}=\{H_I, W_I, C_I\}$ and $C_K$ kernels each of $\{height, width, channels\}=\{H_K, W_K, C_I\}$, the "im2col" technique constructs an "input-patch" matrix, $P$, by unrolling patches from the input into columns. The size of $P$ is $(H_K * W_K * C_I)\times (H_O * W_O)$, where $H_O=(H_I-H_K+2*(\#ZeroPaddings))/Stride+1$ and $W_O=(W_I-W_K+2*(\#ZeroPaddings))/Stride+1$. A "kernel-patch" matrix, $K$, is formed by unrolling each of the $C_K$ kernels of shape $k\times k\times C$ into a row. The size of $K$ is $C_K\times (H_K*W_K*C_I)$. Performing the general matrix multiplication $K\times P$ yields an output matrix, $O$, of size $C_K\times (H_O*W_O)$. Fig. \ref{convex} illustrates a GEMM example.

\begin{figure}[ht!]
\centering 
{\includegraphics[width=0.45\textwidth]{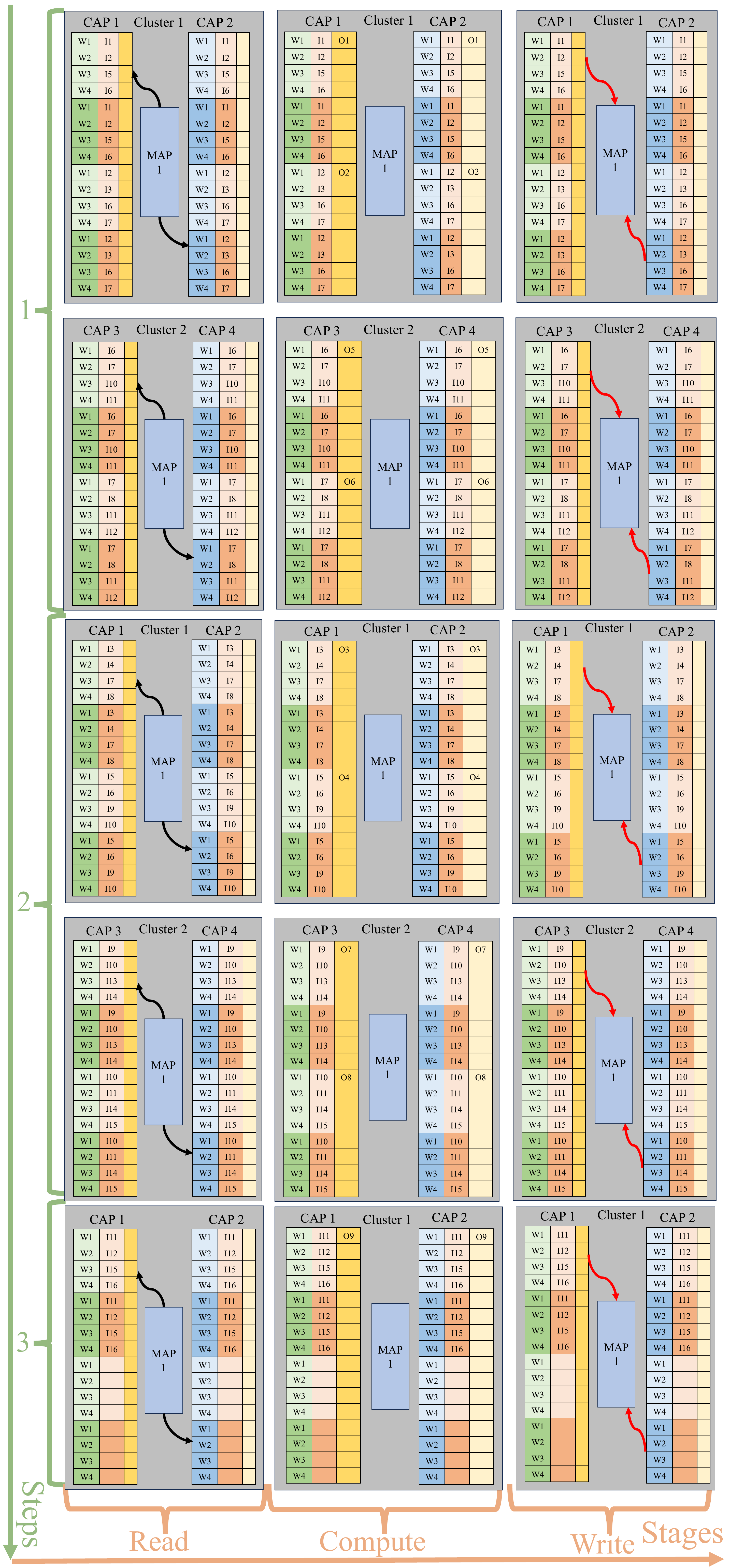}}
\caption{Example showing the mapping of the convolutional layer in Fig. \ref{convex} on a $2\times 2$ BF-IMNA architecture.}
\label{ws-example}
    \vspace{-0.15in}
\end{figure}

\begin{table*}[ht!]
\centering
\caption{Devised run time of functions on APs. $L=$\#words stored in AP, $M=$bitwidth/word, matrices for multiplication are $i\times j$ and $j\times u$ matrices. For pooling, $S=$window size and $K=$\#operations. $J=\log_2(S)$.}
\label{runtime}
\resizebox{1\textwidth}{!}{%

\begin{tabular}{|*{4}{c|}}
\hline
 & \multicolumn{3}{c|}{Runtime} \\ \hline
\diagbox[width=10em]{Function}{AP Type}& \multicolumn{1}{c}{1D AP} & \multicolumn{1}{|c}{2D AP (no segmentation)} & \multicolumn{1}{|c|}{2D AP (segmentation)} \\ \hline
Addition & \multicolumn{1}{c}{$2M+8M+M+1$} & \multicolumn{1}{|c}{$2M+8M+M+1$} & \multicolumn{1}{|c|}{$2M+8M+M+1$}\\
\hline
Multiplication & \multicolumn{1}{c}{$2M+8M^2+2M$} & \multicolumn{1}{|c}{$2M+8M^2+2M$} & \multicolumn{1}{|c|}{$2M+8M^2+2M$}\\
\hline
Reduction & \multicolumn{1}{c}{$2M+\sum^{\log_2(L)}_{q=1}(8(M+q-1))+L-1$} & \multicolumn{1}{|c}{$2M+8M+8(L/2-1)+1$} & \multicolumn{1}{|c|}{$2M+8M+8\log_2(L/2)+1$}\\
\hline
Matrix-Matrix Multiplication & \multicolumn{1}{c}{$2M+8M^2+\sum^{\log_2(j)}_{q=1}(8(2M+q-1))+2(i*u)(j-1)+2M+\log_2(j)$}  & \multicolumn{1}{|c}{$2M+8M^2+8(i*u)(j-1)+2M+\log_2(j)$}  &\multicolumn{1}{|c|}{$2M+8M^2+8\log_2(j)+2M+\log_2(j)$}\\ \hline
ReLU & \multicolumn{1}{c}{$4M+1$} & \multicolumn{1}{|c}{$4M+1$} & \multicolumn{1}{|c|}{$4M+1$}\\ \hline

Max Pooling & \multicolumn{1}{c}{$2M+(8M+2)(\log_2(S))+2K(S/2-1)+M$} & \multicolumn{1}{|c}{$2M+(8M+2)+10K(S/2-1)+M$} & \multicolumn{1}{|c|}{$2M+(8M+2)+(8+2K)\log_2(S/2)+M$} \\ \hline
Average Pooling & \multicolumn{1}{c}{$2M+2K(S/2-1)+\sum^{\log_2(S)}_{q=1}(8(M+q-1))+M$}  & \multicolumn{1}{|c}{$2M+8M+8K(S/2-1)+M$}  &\multicolumn{1}{|c|}{$2M+8M+8\log_2(S/2)+M$}\\
\hline
\end{tabular}
}
\end{table*}

\begin{table*}[ht!]
\centering
\caption{Devised complexities of functions on 1D AP and 2D AP.}
\label{complexities}
\resizebox{1\textwidth}{!}{%
\begin{tabular}{|*{4}{c|}}
\hline
 & \multicolumn{3}{c|}{Complexity} \\ \hline
\diagbox[width=10em]{Function}{AP Type}& \multicolumn{1}{c}{1D AP} & \multicolumn{1}{|c}{2D AP (no segmentation)} & \multicolumn{1}{|c|}{2D AP (segmentation)} \\ \hline
Addition & \multicolumn{1}{c}{$O(M)$} & \multicolumn{1}{|c}{$O(M)$} & \multicolumn{1}{|c|}{$O(M)$}\\
\hline
Multiplication & \multicolumn{1}{c}{$O(M)+O(M^2)$} & \multicolumn{1}{|c}{$O(M)+O(M^2)$} & \multicolumn{1}{|c|}{$O(M)+O(M^2)$}\\
\hline
Reduction & \multicolumn{1}{c}{$O(M)+O(M\log_2(L))+O(L)$} & \multicolumn{1}{|c|}{$O(M)+O(L)$} & \multicolumn{1}{c|}{$O(M)+O(\log_2(L))$}\\ \hline
Matrix-Matrix Multiplication & \multicolumn{1}{c}{$O(M)+O(M^2)+O(M\log_2(j))+O(i*u*j)$} & \multicolumn{1}{|c|}{$O(M)+O(M^2)+O(i*u*j)$}  &\multicolumn{1}{c|}{$O(M)+O(M^2)+O(\log_2(j))$}\\ \hline
ReLU & \multicolumn{1}{c}{$O(M)$} & \multicolumn{1}{|c|}{$O(M)$}&\multicolumn{1}{c|}{$O(M)$} \\ \hline
Max Pooling & \multicolumn{1}{c}{$O(M)+O(M\log_2(S))+O(S*K)$} & \multicolumn{1}{|c|}{$O(M)+O(S*K)$}&\multicolumn{1}{c|}{$O(M)+O(\log_2(S))+O(K\log_2(S))$} \\ \hline

Average Pooling & \multicolumn{1}{c}{$O(M)+O(SK)+O(M\log_2(S))$} & \multicolumn{1}{|c|}{$O(M)+O(SK)$}  &\multicolumn{1}{c|}{$O(M)+O(\log_2(S))$} \\ \hline
\end{tabular}%
}
    \vspace{-0.15in}
\end{table*}

\begin{figure}[hbt!]
\centering
\subfloat[]{\includegraphics[width=0.45\columnwidth]{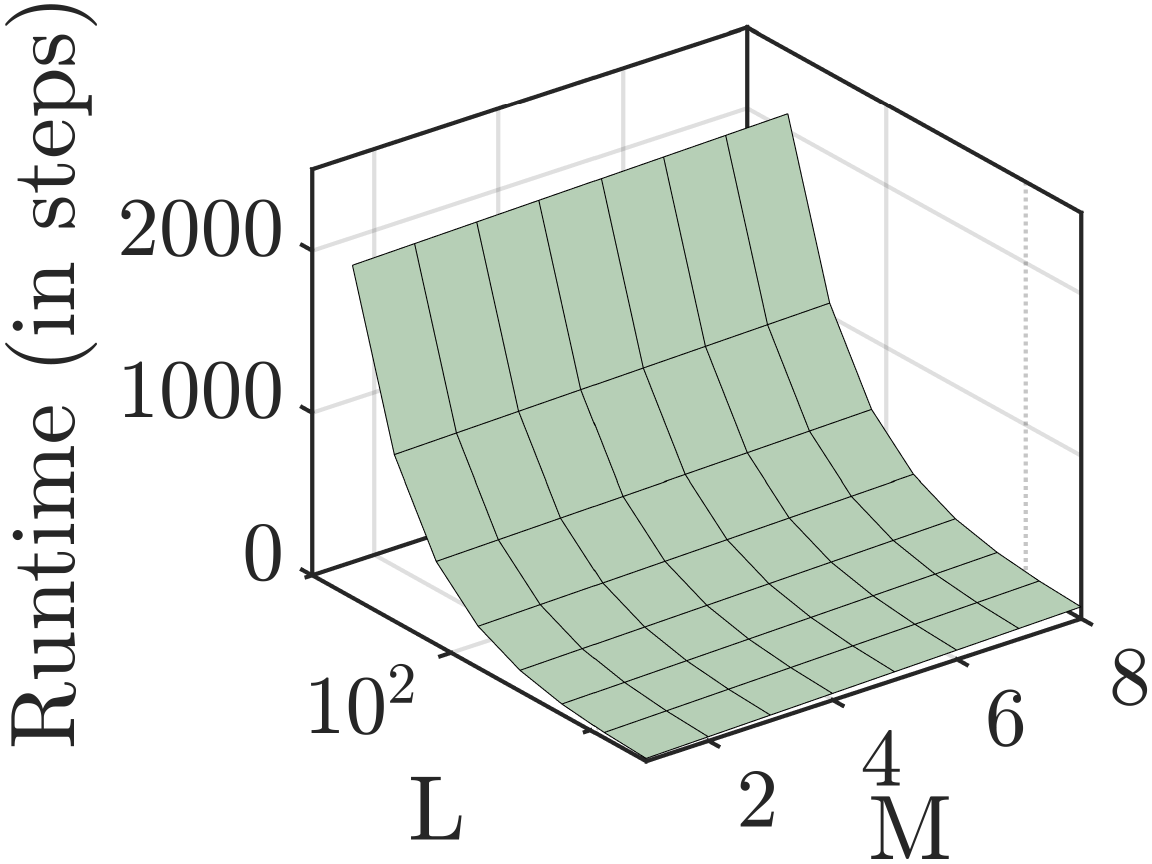}}
\hfil
\subfloat[]{\includegraphics[width=0.45\columnwidth]{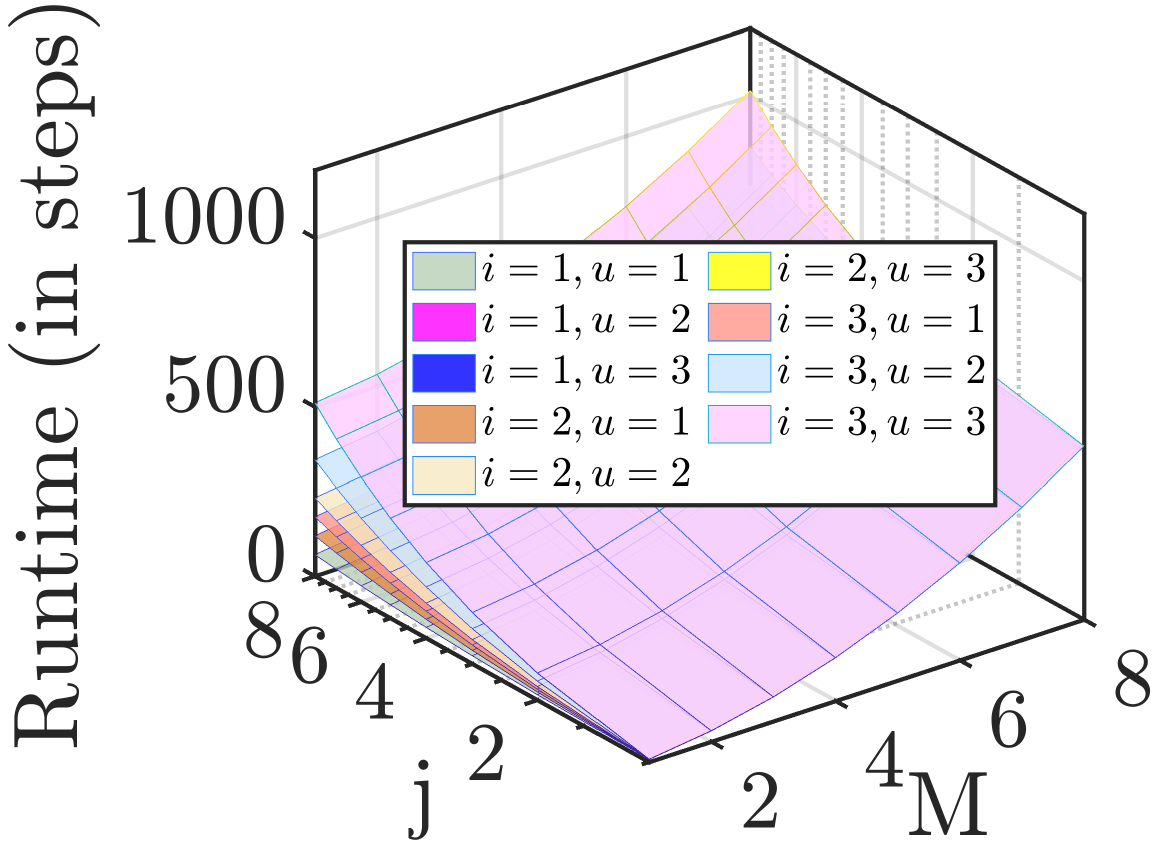}}
\hfil
\subfloat[]{\includegraphics[width=0.43\columnwidth]{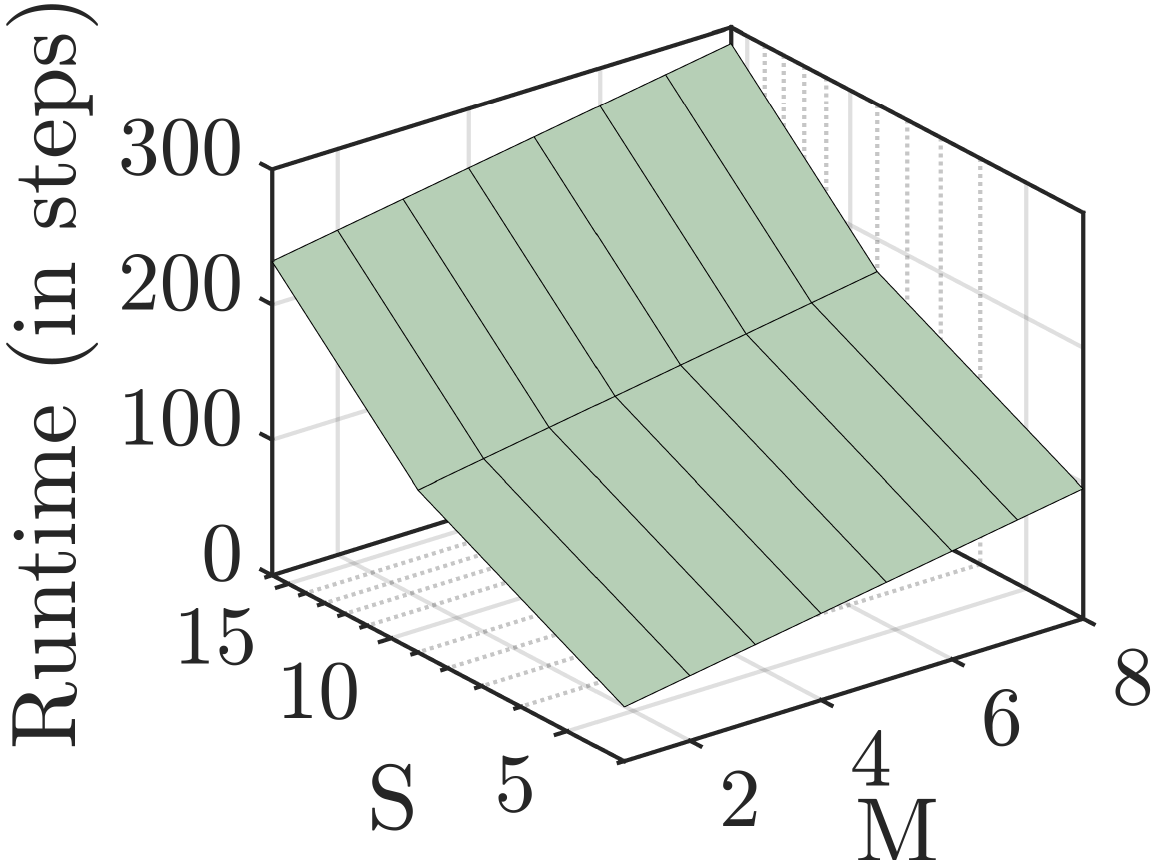}}
\hfil
\subfloat[]{\includegraphics[width=0.43\columnwidth]{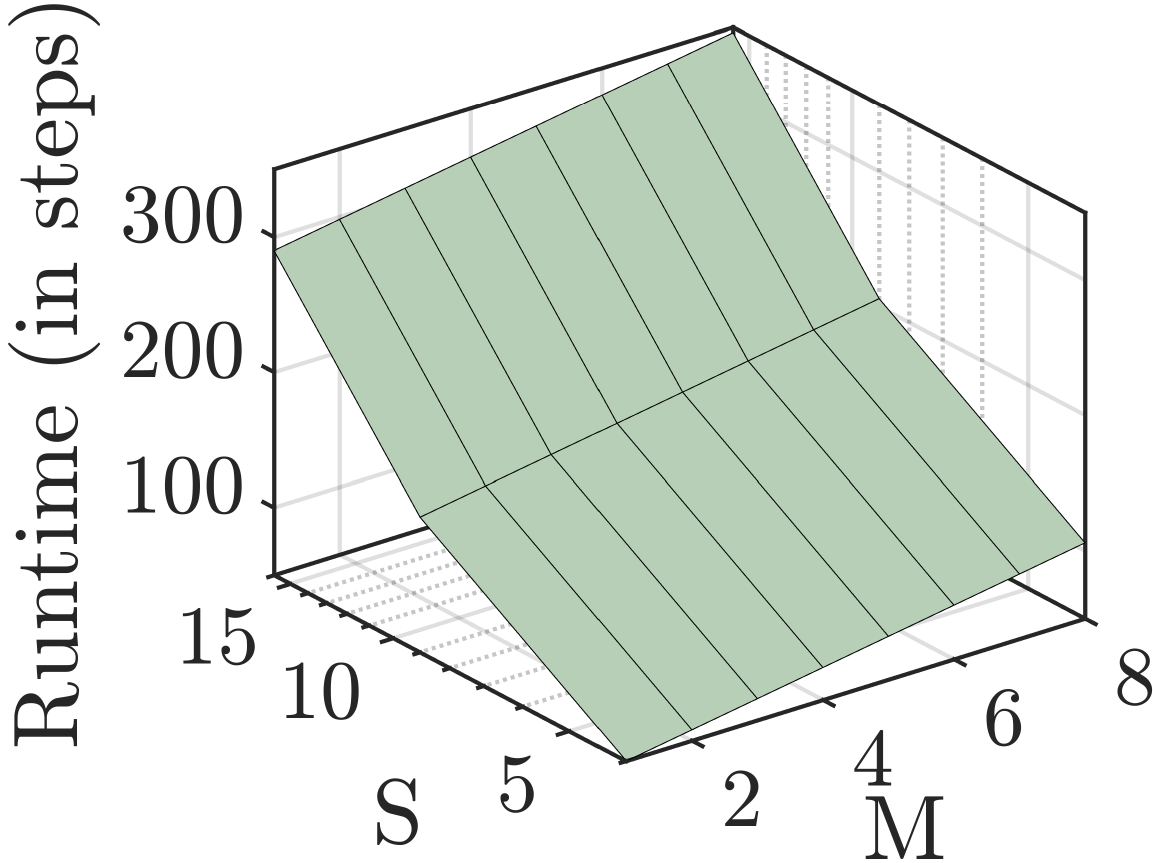}}
\hfil
\subfloat[]{\includegraphics[width=0.30\columnwidth]{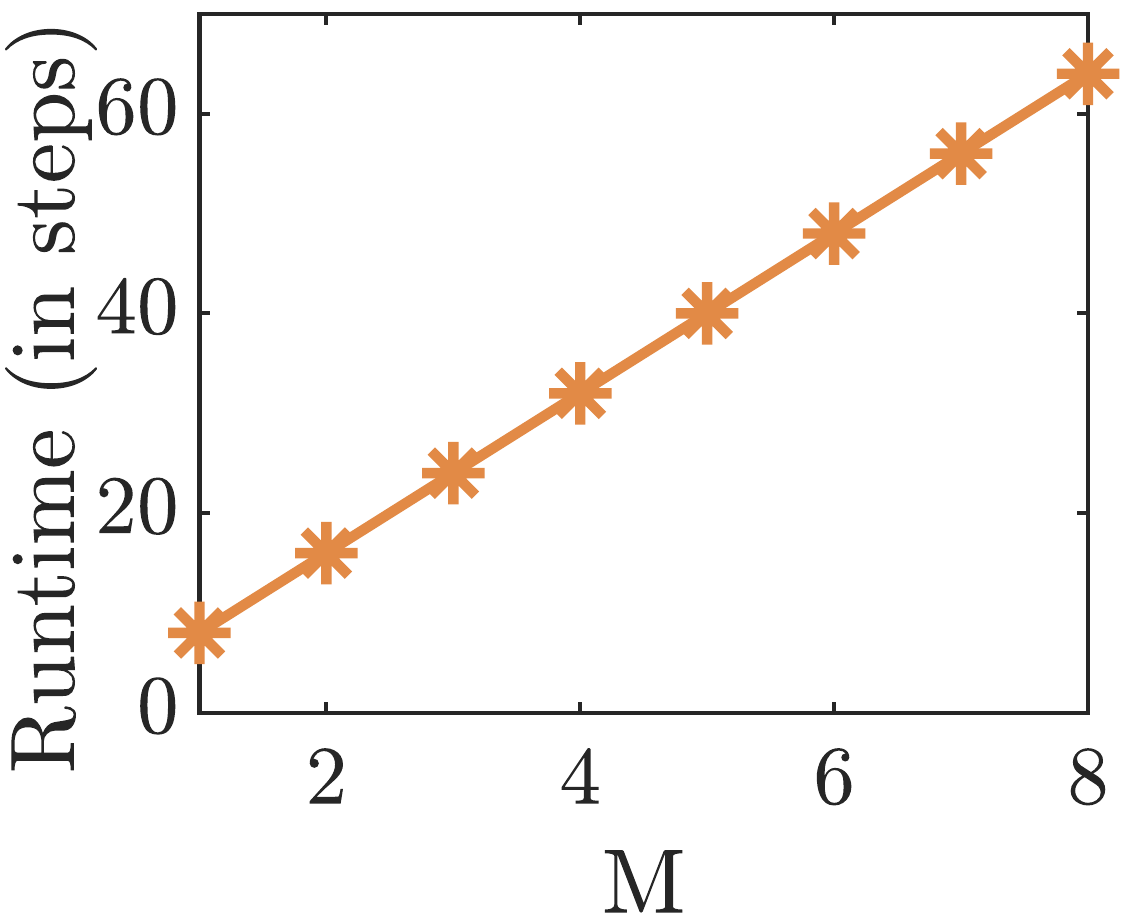}}
\hfil
\subfloat[]{\includegraphics[width=0.34\columnwidth]{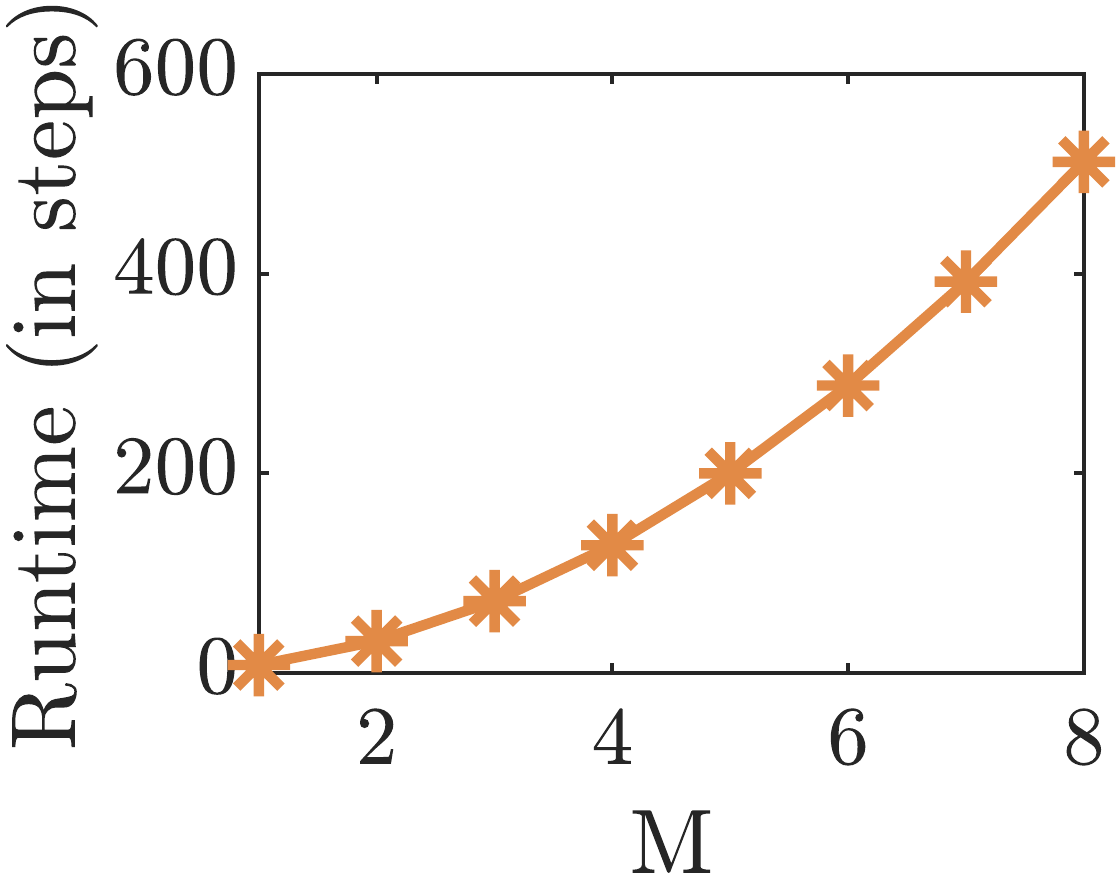}}
\hfil
\subfloat[]{\includegraphics[width=0.34\columnwidth]{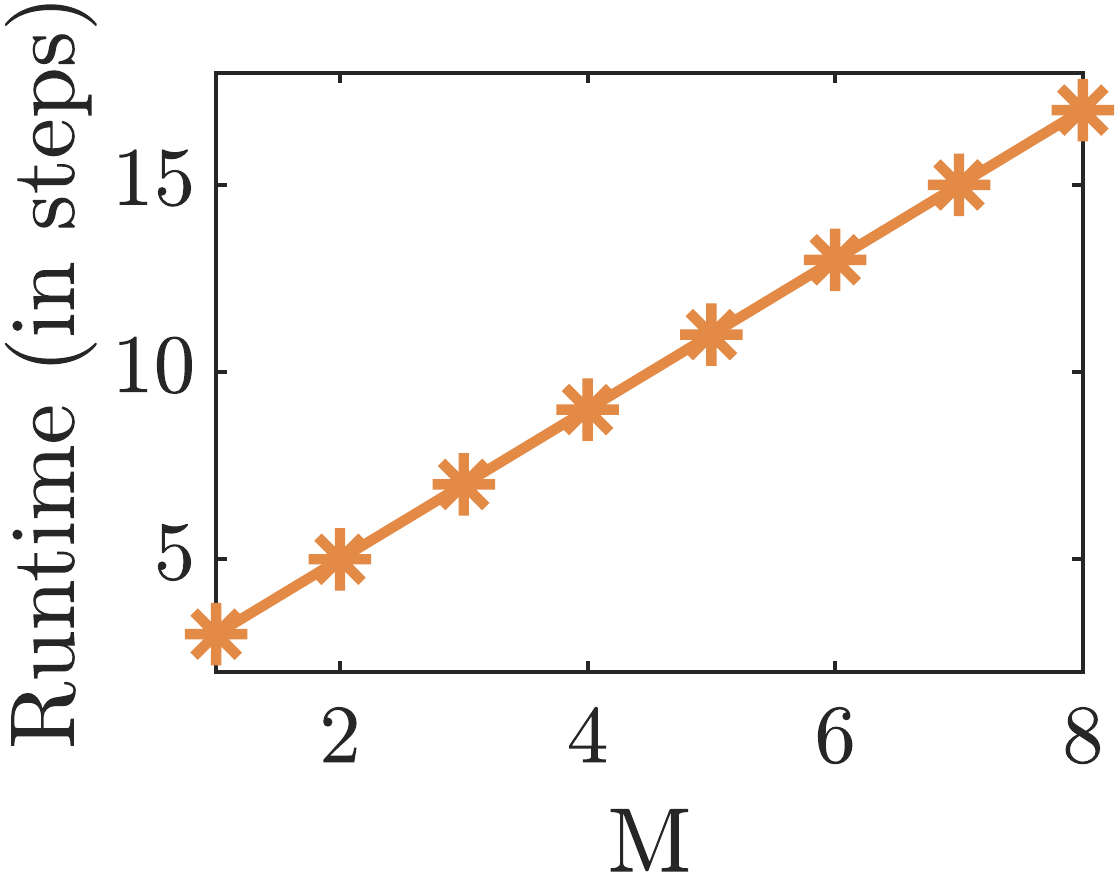}}
\caption{AP Runtime of (a) reduction, (b) matrix-matrix multiplication, (c) average pooling, (d) max pooling, (e) addition, (f) multiplication, and, (g) ReLU.\label{runtimes}}
    \vspace{-0.2in}
\end{figure}

\begin{table}[hb]
    \centering
    \caption{LUT of ReLU. $F_i:$ flag bits in column/row $i$ and $A_i:$ selected column/row. NC = no change.}
    \begin{tabular}{|c|c|c|}
    \hline
        $A_i$/$F_i$ & Pass & Resulting $A_i$ \\ \hline
         10 & NC & 1 \\ \hline
         01 & NC & 0 \\ \hline
         11 & 1st & 0 \\ \hline
         00 & NC & 0 \\ \hline
    \end{tabular}
    \label{lutrelu}
\end{table}

\begin{table}[hb!]
    \centering
    \caption{The LUT of the max pooling function. $F_{i1}$ and $F_{i2}$ are the flag bits in column/row $i$. $A_i$ and $B_i$ are the selected columns/rows in the words of interest. NC = no change and NP = not possible.}
    \scalebox{0.9}{%
    \begin{tabular}{|c|c|c|c|c|}
    \hline
        $A_i$/$B_i$/$F_{i1}$/$F_{i2}$ & Pass & Resulting $B_i$ & Resulting $F_{i1}$ & Resulting $F_{i2}$\\ \hline
         1010 & NP & 0 & 1 & 0 \\ \hline
         0110 & NP & 1 & 1 & 0 \\ \hline
         1110 & NP & 1 & 1 & 0 \\ \hline
         0010 & NP & 0 & 1 & 0 \\ \hline
         1000 & 1st & 1 & 0 & 1 \\ \hline
         0100 & 2nd & 1 & 1 & 1 \\ \hline
         1100 & NC & 1 & 0 & 0 \\ \hline
         0000 & NC & 0 & 0 & 0 \\ \hline
         1011 & NC & 0 & 1 & 1 \\ \hline
         0111 & NC & 1 & 1 & 1 \\ \hline
         1111 & NC & 1 & 1 & 1 \\ \hline
         0011 & NC & 0 & 1 & 1 \\ \hline
         1001 & 3rd & 1 & 0 & 1 \\ \hline
         0101 & 4th & 0 & 0 & 1 \\ \hline
         1101 & NC & 1 & 0 & 1 \\ \hline
         0001 & NC & 0 & 0 & 1 \\ \hline
    \end{tabular}}
    \label{lutmaxpooling}
\end{table}

\section{BF-IMNA}
\subsection{Hardware Configurations}
BF-IMNA comprises cluster(s), each containing multiple CAPs and one MAP as shown in Fig. \ref{arch}. MAPs communicate with both CAPs and off-chip memory via an on- and off-chip interconnect, respectively. We map (based on GEMM) workloads into BF-IMNA. We note that the mapping does not change for different precisions: all precisions map the same way to the hardware, and for lower precision, Most Significant Bits (MSBs) are deactivated for energy savings. We study two configurations of BF-IMNA: maximum parallelism and limited resources, that we elaborate on next.

\noindent\textbf{Maximum Parallelism}: Here, we aim to achieve full spatial dimension computation unrolling, utilizing Infinite Resources (IR) to exploit maximum intra-layer parallelism. In this configuration, a single large cluster houses all required CAPs for computing the largest layer in one step, along with a sufficiently large MAP for streaming inputs to CAPs through an on-chip mesh interconnect. Configuring the accelerator size is based on the dimensions of the convolutional layer with the highest number of multiply-accumulates (MACs). We transfer all weights offline from off-chip memory to CAPs before inference. BF-IMNA assumes a sufficiently large on-chip memory (MAPs) to store the model weights, which avoids off-chip memory accesses that significantly increase energy consumption, particularly in the Limited Resources (LR) setting (see below). The area reported in Table V includes enough on-chip area to store the weights of our largest studied model (VGG16). During inference, input streaming involves moving data from off-chip memory to MAP via off-chip interconnect and then to CAPs via mesh for processing. Between layers, two data movements are necessary: 1) streaming current layer weights and 2) rearranging outputs of one layer to serve as inputs for the next layer. The MAPs act as intermediate storage for rearranging CAP outputs and managing streaming weights to CAPs. To reshape the output of a layer in preparation for the next layer, we perform the following steps: 1- outputs are read in word-sequential mode from CAPs, then 2- transferred over the bus to the MAP, where they are 3- written in word-sequential mode, in consecutive rows. At this point, the outputs of the current layer are reshaped. 4- The reshaped outputs are read from the MAP, 5- transferred over the on-chip bus, and 6- written back to the CAPs in the rearranged format word-sequentially. All reshaping overheads are factored into our results. Additionally, the latency of writing input/weights and intermediate outputs in the MAP is hidden by data transfer through the mesh.

\noindent\textbf{Limited Resources}
IR-based mapping is resource-intensive. A more practical approach restricts the accelerator size to avoid an expensive design. In this configuration, we use $4096$ CAPs, organized as an $8\times 8$ CAPs within each cluster of the $8\times 8$ clusters. We chose these numbers because our experiments have shown such a configuration achieves nearly 100\% hardware utilization while striking a balance with time-folding overheads. Fig. \ref{arch} illustrates a smaller version of this hardware configuration, where clusters operate independently and in parallel, eliminating the need for interconnect between them. Although an interconnect is necessary between off-chip memory and MAPs inside each cluster, its cost is omitted from our results section, as there is sufficient on-chip memory to program all weights in MAPs offline. Due to insufficient computational resources for computing an entire layer output in a single step, we use a weight stationary mapping, performing GEMM over multiple time steps. The weight matrix streams from MAPs onto CAPs via the mesh and remains stationary in time as new columns from the input matrix are introduced at every step. Particularly, if the weight kernel $K_i$ in layer $i$ exceeds CAP capacity, we fold the mapping in time. Each cluster stores a copy of $K_i$ across its APs and computes different columns of the output matrix. During inference, for each layer, each cluster receives different columns from the input matrix $P_i$, broadcasted to CAPs via the corresponding MAP at every step until output computation finishes. Between layers, the MAP serves as intermediate storage for rearranging output into inputs for the next layer and streaming the new weight kernel $K_{i+1}$ as detailed above (in Maximum Parallelism). Similar intra- and inter-layer latency-hiding techniques are employed as described in the IR-based configuration. Fig. \ref{ws-example} illustrates an intra-layer convolutional operation mapping onto BF-IMNA with Limited Resources (LR), assuming a simple convolutional layer (Fig. \ref{convex}). Each step comprises multiple "time" steps, with three stages: 1) Read, where inputs are read from the MAP and written in the CAPs in each cluster, 2) Compute, where CAPs perform convolution, and 3) Write, where outputs are written in arranged form in the MAP from the CAPs in each cluster. Cluster 2 is excluded from step 3 as it is unused.

\subsection{AP Modeling}

We formulate the runtime models of performing some functionalities on the 1D and 2D AP (with and without segmentation). We divide the models into three categories: micro functions, macro functions, and CNN functions. We elaborate on the runtimes of those models below and provide the summaries in Tables \ref{runtime} and \ref{complexities}. We assume that we store 2 words, each with a precision of $M$ in each row of the AP (except for ReLU functionality where the AP is assumed to store all the words vertically). We also assume that there is a total of $L$ words stored in the AP.

\subsubsection{Micro Functions}
Micro functions comprise preliminary operations whose models will be used to build the models of macro functions (below). We focus on three such functions: addition, multiplication, and reduction.

\noindent\textit{Addition:} Given two vectors $A=[a_1, a_2, ..., a_{L/2}]$ and $B=[b_1, b_2, ..., b_{L/2}]$, performing the in-place addition $A+B=B$ on the AP requires performing the following steps: 1-) populating the AP with data by performing $2M$ column writes: bit-sequential mode is assumed for purposes of modeling as the number of rows is assumed to exceed the number of columns and hence a word-sequential write mode is inefficient. 2-) Performing the addition by following the series of compare and write as dictated by the corresponding LUT (can be found in \cite{yantir2018efficient}), and 3-) reading out the resulting B vector in bit-sequential mode ($M+1$ column reads). We do not require adding the results between rows, hence only the horizontal mode is used in the case of 2D AP (with and without segmentation). In this case, the average runtime of the in-place addition for 1D and 2D AP is identical.
\begin{equation}
   rt_{add}=(2M)_{write}+(4M)_{compare}+(4M)_{write}+(M+1)_{read} 
\end{equation}
As there are four passes in the truth table, per each pair of columns, each pass will comprise one compare phase and one write phase in the AP. The write phase is a result of one or more rows matching during the compare, and only one is needed since we can write the whole AP in one shot, regardless of how many rows have matched during the compare phase of that pass.

\noindent\textit{Multiplication:} Similar to addition, out-of-place multiplication ($A*B=C$) comprises the three steps mentioned above (with the second step being performing multiplication by following the series of compare and write as dictated by the corresponding LUT). The average runtime of the out-of-place multiplication for 1D and 2D AP is as follows.
\begin{align*}
rt_{multiply}=&(2M)_{write}+(4M^2)_{compare}+
   (4M^2)_{write}+\\&(2M)_{read} \numberthis
\end{align*}
We note that we do $2M$ reads as the result's precision is double that of the operands.

\noindent\textit{Reduction:} As for reduction, given a vector $A=[a_1, a_2, .. a_{L}]$, the reduction operation adds all of the elements of the vector, i.e. $a_1+a_2+...+a_{L}$. For the 1D AP, this comprises four main steps: 1-) populating AP with data in a bit-sequential mode ($2M$ column writes) where each pair of consecutive elements in the vector are stored in one row (number of rows in AP is still $L/2$), 2-) in-place addition in horizontal mode is performed on elements in the AP as explained in \cite{yantir2018efficient} to get the sums {$a_1+a_2=a_2$, $a_3+a_4=a_4$, ..., $a_{L-1}+a_{L}=a_{L}$} (henceforth we refer to one in-place addition in horizontal mode as performing four compares and four writes per a pair of columns for all column pairs in the AP). 3-) Then the resulting words in ${2^{nd} Row, 3^{rd} Row, ..., {L/2}^{th} Row}$ are transferred to be stored next to the resulting words in ${1^{st} Row, 2^{nd} Row, ...., {L/2-1}^{th} Row}$ respectively. The words are transferred sequentially (i.e. 1 word transferred at a time in word-sequential mode, where one transfer comprises one read and one write). The second and third step (horizontal in-place addition and data transfer) are repeated to complete the reduction (in total in-place addition is done $\log_2(L)$ times and data is transferred $(L/2-1)$ times. We note that after each time step 2- is done, the width of the resulting word increases by one bit). 4-) The final result of the reduction (one word) is read using word-sequential mode. In the case of 2D AP (with and without segmentation), reduction also comprises four steps: 1-) populating the AP with data, 2-) performing in-place addition in horizontal mode, 3-) performing in-place addition in vertical mode. Vertical mode means we perform the addition between pairs of rows, hence not needing to constantly transfer data (unlike in 1D AP) for consecutive horizontal mode in-place additions. For 2D AP without segmentation, one pair of rows is considered at a time, while with segmentation, all pairs of rows are operated in parallel. Step 3-) is repeated as much as needed to finish the reduction (i.e. $(L/2-1)$ times, and $\log_2(L/2)$ times in the case of 2D AP and 2D AP with segmentation respectively). Note that the vertical mode reduction in the 2D AP with segmentation is performed in parallel (similar to having a reduction tree next to the 1D AP). 4-) The final result of the reduction (one word) is read using word-sequential mode. Below are the run times of reduction in cases of 1D, 2D AP, and 2D AP with segmentation respectively.
\begin{align*}
   &rt_{reduce1D}=(2M)_{write}+(1_{read}+1_{write})*(L/2-1) \\
   &+\sum^{\log_2(L)}_{q=1}((4(M+q-1))_{compare}+(4(M+q-1))_{write})\\&+1_{read} \numberthis
\end{align*}
\begin{align*}
   rt_{reduce2D}=(2M)_{write}+(4M)_{compare}+(4M)_{write}\\
   +(L/2-1)*((4)_{compare}+(4)_{write})+1_{read} \numberthis
\end{align*}
\begin{align*}
   &rt_{reduce2DwSeg}=(2M)_{write}+(4M)_{compare}+(4M)_{write}\\
   &+\log_2(L/2)*((4)_{compare}+(4)_{write})+1_{read} \numberthis
\end{align*}
Note that in the case of 2D AP with segmentation, we need to have $L/4$ rows of $Carry$ to perform the reductions in parallel across all row pairs in the vertical mode.

\subsubsection{Macro Functions} We define a macro function as one that relies on more than one micro function. Matrix-matrix multiplication and dot product are macro functions of interest.

\noindent\textit{Matrix-matrix Multiplication:} Matrix-matrix multiplication comprises multiplying an $i\times j$ matrix by an $j\times u$ matrix. As dot product is a special case of matrix-matrix multiplication where $i=u=1$, we will henceforth focus on matrix-matrix multiplication. Still assuming that we store 2 words per AP row to perform the computation, the number of rows needed in the AP to perform the matrix-matrix multiplication is in this case $i*j*u+1$ or $i*j*u+(i*j*u)/2$ (the additional row(s) is(are) for carry row(s) used in every row pair computation in 2D AP without(with) segmentation). In the case of 1D AP, matrix-matrix multiplication comprises five steps: 1-) populating the AP with data in bit-sequential mode ($2M$ column writes), 2-) performing out-of-place multiplication in the horizontal mode, and 3-) transferring the words in word-sequential mode. 4-) Performing in-place addition in the horizontal mode. Steps 3-) and 4-) are repeated to complete the computation (in total there are $\log_2(j)$ in-place additions and $(i*u)*(j-1)$ transfers). 5-) The resulting values ($i*u$ of them) are read out in bit-sequential mode ($(2M+\log_2(j))$ reads). As for the 2D AP, matrix-matrix multiplication comprises four steps: 1-) populating data in AP, 2-) performing out-of-place multiplication in horizontal mode, 3-) performing a set of in-place additions in the vertical mode: depending on the sizes of the matrices and type of 2D AP, several reductions are needed ($(i*u)*(j-1)$ times in 2D AP without segmentation and $\log_2(j)$ times in 2D AP with segmentation). For the 2D AP with segmentation, the reductions across all rows are carried out in parallel rather than each pair of rows at a time. Finally, 4-) the resulting values ($i*u$ of them) are read out in bit-sequential mode ($(2M+\log_2(j))$ reads). Below are the run times of matrix-matrix multiplication in cases of 1D, 2D AP, and 2D AP with segmentation respectively.
\begin{align*}
   &rt_{matmat1D}=(2M)_{write}+(4M^2)_{compare}+(4M^2)_{write}\\
   &+\sum^{\log_2(j)}_{q=1}((4(2M+q-1))_{compare}+(4(2M+q-1))_{write})\\
   &+(1_{read}+1_{write})*(i*u)*(j-1)+(2M+\log_2(j))_{read} \numberthis
\end{align*}
\begin{align*}
   &rt_{matmat2D}=(2M)_{write}+(4M^2)_{compare}+(4M^2)_{write}\\
   &+(4_{compare}+4_{write})(i*u)*(j-1)+(2M+\log_2(j))_{read} \numberthis
\end{align*}
\begin{align*}
   &rt_{matmat2DwSeg}=(2M)_{write}+(4M^2)_{compare}+(4M^2)_{write}\\
   &+\log_2(j)*(4_{compare}+4_{write})+(2M+\log_2(j))_{read} \numberthis
\end{align*}

Notice how the final result has a $2M+\log_2(j)$ bitwidth. This is because the multiplication doubles the bitwidth then each reduction adds an extra bit to the bitwidth.

\subsubsection{CNN Functions} 
CNN functions are specific to CNN inference: average pooling, max pooling, and ReLU.

\noindent\textit{Average Pooling:} Assuming a window size of $S$ and $K$ pooling operations, average pooling is performed on vectors $A=[a_1, a_2, ..., a_{L/2}]$ and $B=[b_1, b_2, ..., b_{L/2}]$. For demonstration, and without loss of generality, assume that $S=K=4$. In that case, there are $L=S*K=16$ elements that will be stored in the AP (i.e. $(S*K)/2=8$ rows). Average pooling in 1D AP comprises four main steps: 1-) writing bit-sequentially in the AP ($2M$ column writes), 2-) performing the following in-place additions in horizontal mode and in parallel (we are demonstrating on the word-granularity level) \{$a_1+b_1=b_1$, $a_2+b_2=b_2$, $a_3+b_3=b_3$, $a_4+b_4=b_4$, $a_5+b_5=b_5$, $a_6+b_6=b_6$, $a_7+b_7=b_7$, $a_8+b_8=b_8$\} by following the LUT of in-place addition (four compares and four writes per pair of columns selected). 3-) $b_2$/$b_4$/$b_6$/$b_8$ (the result of the 2nd/4th/6th/8th-row additions) are sequentially transferred (word by word) to replace words $a_1$/$a_3$/$a_5$/$a_7$ in the 1st/3rd/5th/7th rows, where each transfer is one read and one write. Steps 2-) and 3-) are repeated as needed (i.e. there are $K*(S/2-1)=4$ transfers in this case and in total $\log_2(S)=2$ in-place additions). 4-) To mimic dividing by $J=\log_2(S)=2$ (or shifting by 2 to the right), the four outputs are read bit-sequentially starting from the $Jth+1$ least significant bit (i.e. $M-J$ reads). For the 2D AP, average pooling comprises four steps: 1-) populating AP with data, 2-) performing an in-place addition in horizontal mode, 3-) a second and final (in this example) in-place addition is performed in the vertical direction to produce four results. In other examples, 3-) is repeated as needed to produce the final results ($K*(S/2-1)$ times and $\log_2(S/2)$ times in 2D AP and 2D AP with segmentation respectively). 4-) To mimic dividing by $J=\log_2(S)=2$ (or shifting by 2 to the right), the four outputs are read bit-sequentially starting from the $Jth+1$ least significant bit (i.e. $M+\log_2(S)-J=M$ reads). Below are the run times of average pooling.    
\begin{align*}
   &rt_{avpool1D}=(2M)_{write}\\
   &+\sum^{\log_2(S)}_{q=1}((4(M+q-1))_{compare}+(4(M+q-1))_{write})\\
   &+(1_{read}+1_{write})*K*(S/2-1)+(M)_{read} \numberthis
\end{align*}
\begin{align*}
   &rt_{avpool2D}=(2M)_{write}+(4M)_{compare}+(4M)_{write}\\
   &+(4_{compare}+4_{write})K*(S/2-1)+(M)_{read} \numberthis
\end{align*}
\begin{align*}
   &rt_{avpool2DwSeg}=(2M)_{write}+(4M)_{compare}+(4M)_{write}\\
   &+\log_2(S/2)*(4_{compare}+4_{write})+(M)_{read} \numberthis
\end{align*} 
Note that in the case of 2D AP with segmentation, we need to have $L/4$ rows of $Carry$ to perform the reductions in parallel across all row pairs in the vertical mode.

\noindent\textit{Max Pooling:} For max pooling, given a window size of $S$, $K$ pooling operations, and two vectors $A=[a_1, a_2, ..., a_{L/2}]$ and $B=[b_1, b_2, ..., b_{L/2}]$, max pooling in the 1D AP is performed by following five steps: 1-) populating AP bit-sequentially (in $2M$ column writing steps), 2-) Performing the sequences of compare/write of the corresponding LUT that we have formulated in Table \ref{lutmaxpooling} once in the horizontal direction (again here we refer to the sequence of all compares and writes applied to all column pairs as one in-place max pooling operation), and 3-) Resetting the flag bits (two additional columns written). 4-) Transfer the words (one transfer is one read and one write) in word-sequential mode. Steps 2-), 3-), and 4-) are repeated as needed (words are transferred $K*(S/2-1)$ times and in-place max pooling is applied $\log_2(S)-1$ times). 5-) The resulting words are read in bit sequential mode ($M$ reads). In 2D AP (with and without segmentation), max pooling is performed by 1-) populating AP with data, 2-) performing the max pooling operation in horizontal mode, 3-) resetting the flags, 4-) performing max pooling in the vertical mode. Steps 2-), 3-), and 4-) are repeated as needed ($K*(S/2-1)$ times and $\log_2(S/2)$ in cases of 2D AP and 2D AP with segmentation respectively). Finally, 5-) the resulting words are read in bit sequential mode. Below are the run times of max pooling for 1D, 2D AP, and 2D AP with segmentation respectively.   

\begin{align*}
rt_{maxpool1D}=&(2M)_{write}+\\
   &\log_2(S)*((4M)_{compare}+(4M)_{write}+2_{write})\\
   &+(1_{read}+1_{write})*K*(S/2-1)+M_{read} \numberthis
\end{align*}
\begin{align*}
   rt_{maxpool2D}=&(2M)_{write}+(4M)_{compare}+(4M)_{write}\\
   &+K*(S/2-1)*(4_{compare}+4_{write}+2_{write})\\&
   +M_{read} +2_{write}\numberthis
\end{align*}
\begin{align*}
   rt_{maxpool2DwSeg}=&(2M)_{write}+(4M)_{compare}+(4M)_{write}+\\
   & \log_2(S/2)*(4_{compare}+4_{write}+2K_{write})\\&
   +M_{read}+2_{write} \numberthis
\end{align*} 
Note that in the case of 2D AP with segmentation, we need to have $L/4$ rows of $Flag_1$ and $Flag_2$ to perform the LUT passes in parallel across all row pairs in the vertical mode.

\noindent\textit{ReLU:} For ReLU, we assume a vector, $V=[n_1, n_2, .. n_{L}]$ (each element in the vector has a precision of $M$), representing neuron outputs before activation, and after writing bit-sequentially in the AP (in $2M$ writing steps), we store the MSBs of words in the flag bits $F$ and reset the MSB (i.e. two writes and one read). ReLU operation is performed on the pair of selected columns (starting from second MSB and flag and going through all remaining column pairs till the LSB and flag pair) by following the LUT we have constructed in Table \ref{lutrelu}. The final $L$ results are read in a bit sequential mode ($M$ reads).
ReLu's run time is the same for the 1D and 2D AP (with and without segmentation), and it is as follows:
\begin{align*}
rt_{relu}=&M_{write}+(2_{write}+1_{read})+(M-1)_{compare}+\\
&(M-1)_{write}+M_{read}  \numberthis  
\end{align*}

\noindent\textbf{Comments:} 2D AP with segmentation has less time complexity compared to the 1D AP and 2D AP without segmentation, as segmentation allows parallelism in the vertical mode. We still note a good improvement of the 2D AP over the 1D AP, especially when reduction is involved in the operation (i.e. for matrix-matrix multiplication). We note that for our accelerator design, we have assumed a 2D AP without segmentation to favor programmability, generality, and fewer duplicate peripherals. Fig. \ref{runtimes} shows the different run times of micro/macro/CNN functionalities. We will refer to those figures in explaining our results in Section V (see below).

\begin{table}[hbt!]
    \begin{center}
    \caption{BF-IMNA parameters.}
    \resizebox{0.8\columnwidth}{!}{
    \begin{tabular}{|c|c|c|}
    \hline 
    \textbf{Component} & \textbf{Parameter} & \textbf{Spec} 
 \\
    \hline 
      & Technology Node & $16nm$ \\
     Chip & Supported Bitwidth & Up to 8 \\
      & Total Area & $137.45mm^2$\\
      & Frequency & 1GHz \\
    \hline 
    Clusters & Number & $8\times8$ \\
    \hline 
    CAPs/cluster & Number & $8\times8$  \\
     & Size & $4800\times (2*8)$ \\
    \hline 
    MAPs/cluster & Number & 1  \\
     & Size & $4800\times (2*8)$ \\
    \hline
    & Type & Mesh\\
    On-chip interconnect & Average \#Hops & 3.815 \\
    & Frequency & 500MHz \\
    & \#Bits/transfer & 1024 \\
    \hline
    \end{tabular}
    \label{2DAP_NNAparam}
        }
    \end{center}
    \vspace{-0.15in}
\end{table}

    
    

\begin{table}[!t]
\centering
\caption{$16nm$ predictive technology model parameters used for modeling operations of AP cells.}
\label{parameters}
\resizebox{0.8\columnwidth}{!}{
\begin{tabular}{|c|c|c|}
\hline
\textbf{Parameter} & \textbf{Definition} & \textbf{Value} \\ \hline
$E_{wS}$ & SRAM Write Energy & $0.24fJ$ \\ \hline
$E_{wR}$ & ReRAM Write Energy & $21.7pJ$ \\ \hline
$R_{LRS}$ & ReRAM Low Resistance State & $5k\Omega$  \\ \hline
$R_{HRS}$ & ReRAM High Resistance State & $2.5M\Omega$ \\ \hline
$R_{ON}$ & ON Transistor Resistance & $15k\Omega$  \\ \hline
$R_{OFF}$ & OFF Transistor Resistance & $24.25M\Omega$  \\ \hline
$C_{in}$ & Sensing Capacitance & $50fF$  \\ \hline
$V_{DD}$ & Supply Voltage & 1V  \\ \hline
\end{tabular}
}
\end{table}

\section{Hardware Setup and Benchmarks}
We used Python to emulate the AP functionally executing the micro/macro/CNN-functions. A microbenchmark, consisting of random vectors/matrices, was used to validate the proposed mathematical models in Section III. Then, we implemented BF-IMNA's architecture using a Python-based in-house simulator. Our in-house simulator is designed to estimate the performance of  digital AP modules, and thus we do not anticipate any loss in accuracy or operational failures in hardware. Hence, we do not validate accuracy. Also, the software accuracy of different mixed precision workloads with different average precisions is reported in the recent mixed precision survey \cite{rakka2022mixed}.

Given a CNN model, batch size, and AP size (for LR-based mapping), the simulator maps the model layer-by-layer to AP structures using user-chosen configurations: IR/LR. In the case of LR configuration, specs are summarized in Table \ref{2DAP_NNAparam}. Our simulator then estimates performance metrics (energy, latency, $GOPS$, $GOPS/W$, and $GOPS/W/mm^2$) for accelerating end-to-end inference, relying on the complexity models. We followed the same methodology in \cite{yantir2018two,yantir2018efficient,rakka2020design, rakka2023dt2cam}, calibrating our simulator by SPICE simulations with 16nm PTM models. The relevant parameters are shown in Table \ref{parameters}. Additionally, the energy per transfer per $mm$ for the mesh interconnect is sourced from \cite{dally2020domain}, assuming the interconnect operates at half the frequency of the CAPs/MAPs (i.e., at $500MHz$). Using our simulator, we map AlexNet, VGG16, and ResNet50 \cite{simonyan2014very, krizhevsky2017imagenet, he2016deep} on BF-IMNA. We assess ImageNet inference performance with a batch size of 1.

We carry out several experiments where we test the technology: SRAM vs RRAM, the impact of mixed-precision configuration on BF-IMNA's performance, and the impact of voltage scaling.  For our bit fluidity experiment, we rely on HAWQ-V3's \cite{yao2021hawq} precision configurations and map those onto BF-IMNA 
to simulate a layer-wise mixed-precision inference on ResNet18. Without loss of generality, we specifically choose HAWQ-V3's framework amongst the mixed-precision frameworks presented in \cite{rakka2022mixed} to demonstrate BF-IMNA's bit fluidity because the authors were gracious enough to provide the layer-wise precisions yielded by the optimization algorithm. That way, we can directly estimate BF-IMNA 
with that precision mix without having to re-implement the optimization ourselves. We note however that BF-IMNA can be utilized to accelerate any mixed-precision framework. To compare with SOTA accelerators, we report peak performance metrics assuming a fixed-precision for BF-IMNA for fairness.


\begin{figure}[t!]
    \centering
{\includegraphics[width=0.8\columnwidth]{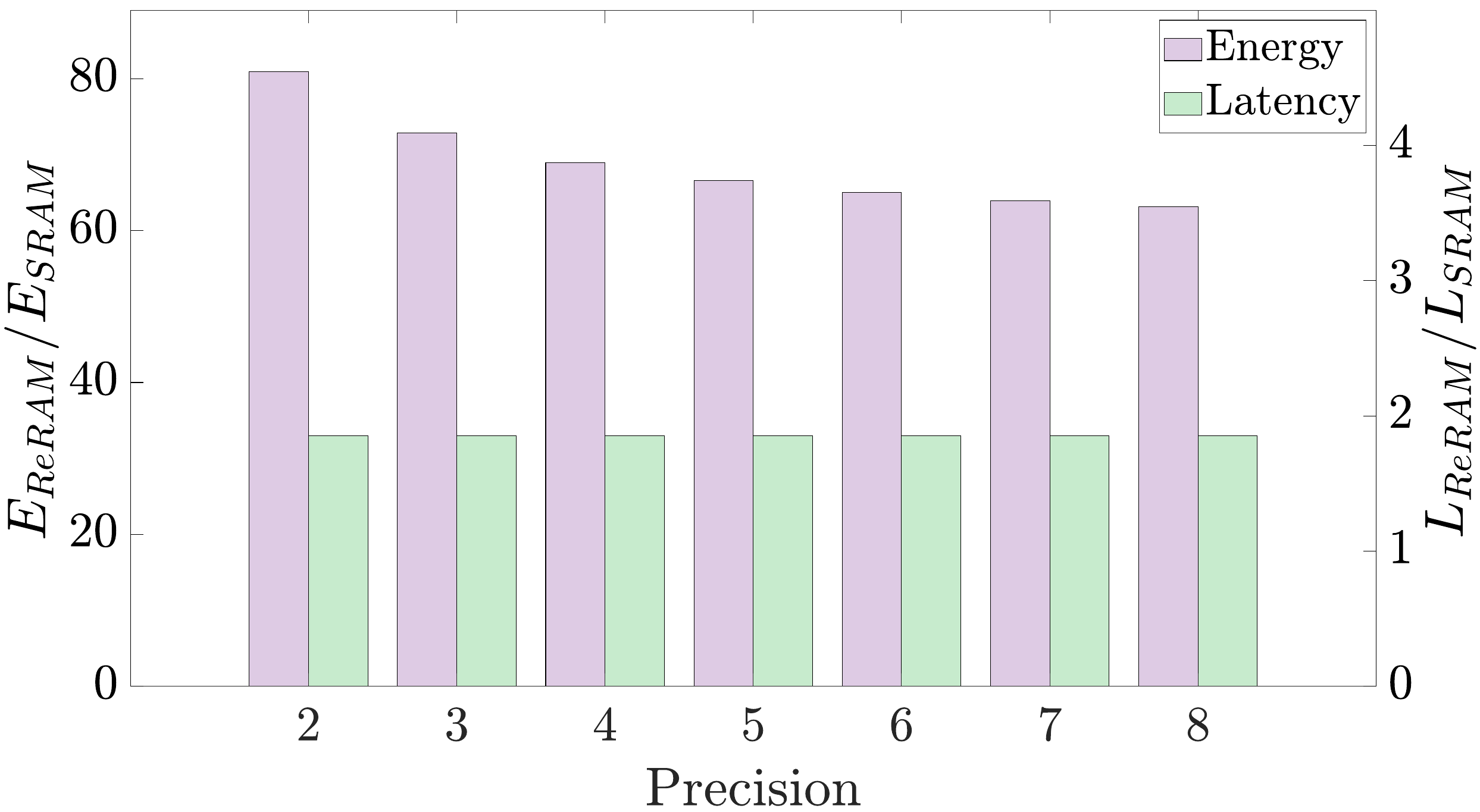}}
\caption{ReRAM/SRAM energy and latency ratios for different precisions with full-fledged inference on VGG16.}\label{rram-sram-ratio}
    \vspace{-0.15in}
\end{figure}

\begin{figure*}[hbt!]
\centering
\subfloat[]{\includegraphics[width=0.333\textwidth]
{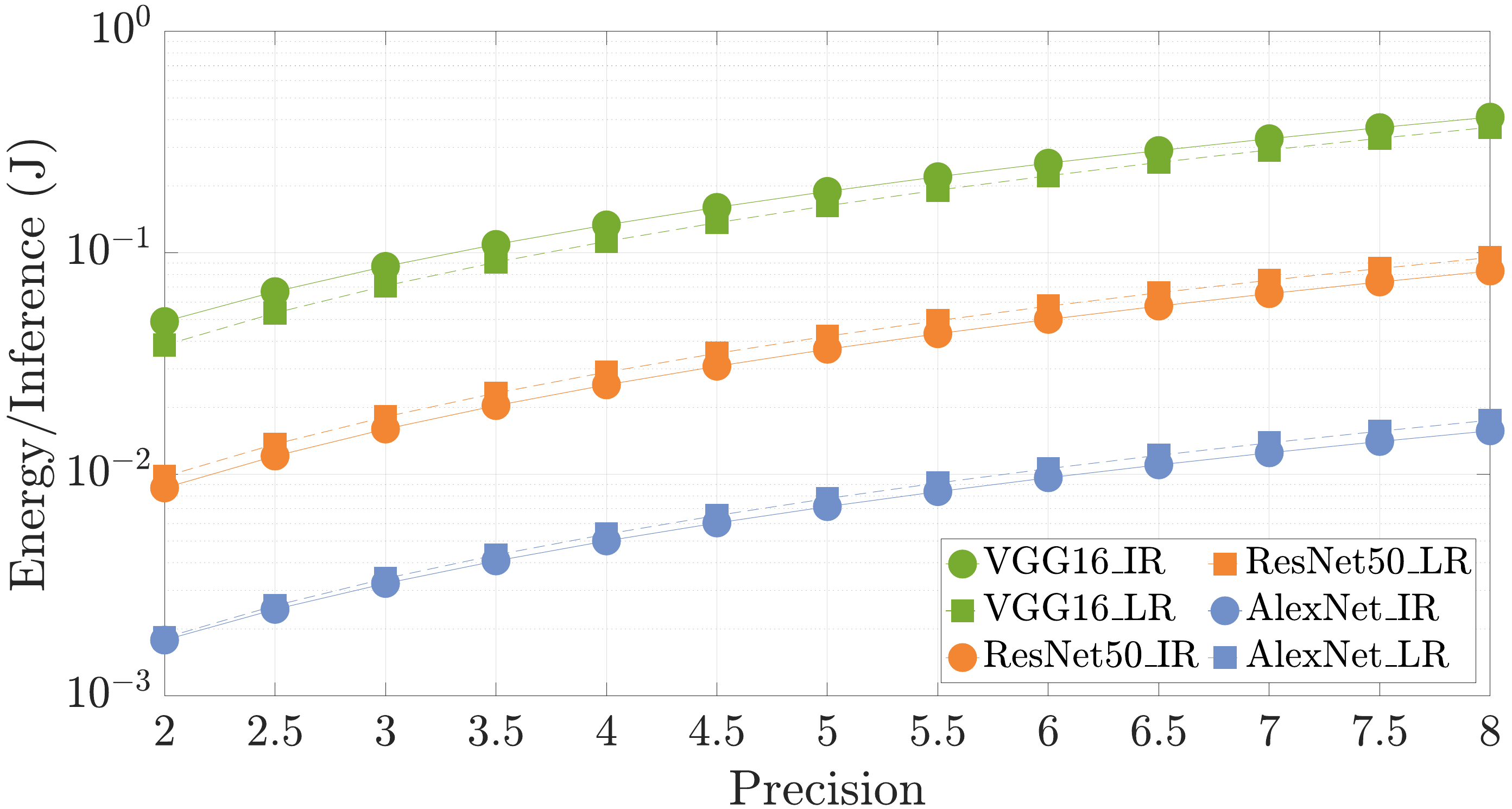}}
\hfil
\subfloat[]{\includegraphics[width=0.333\textwidth]
{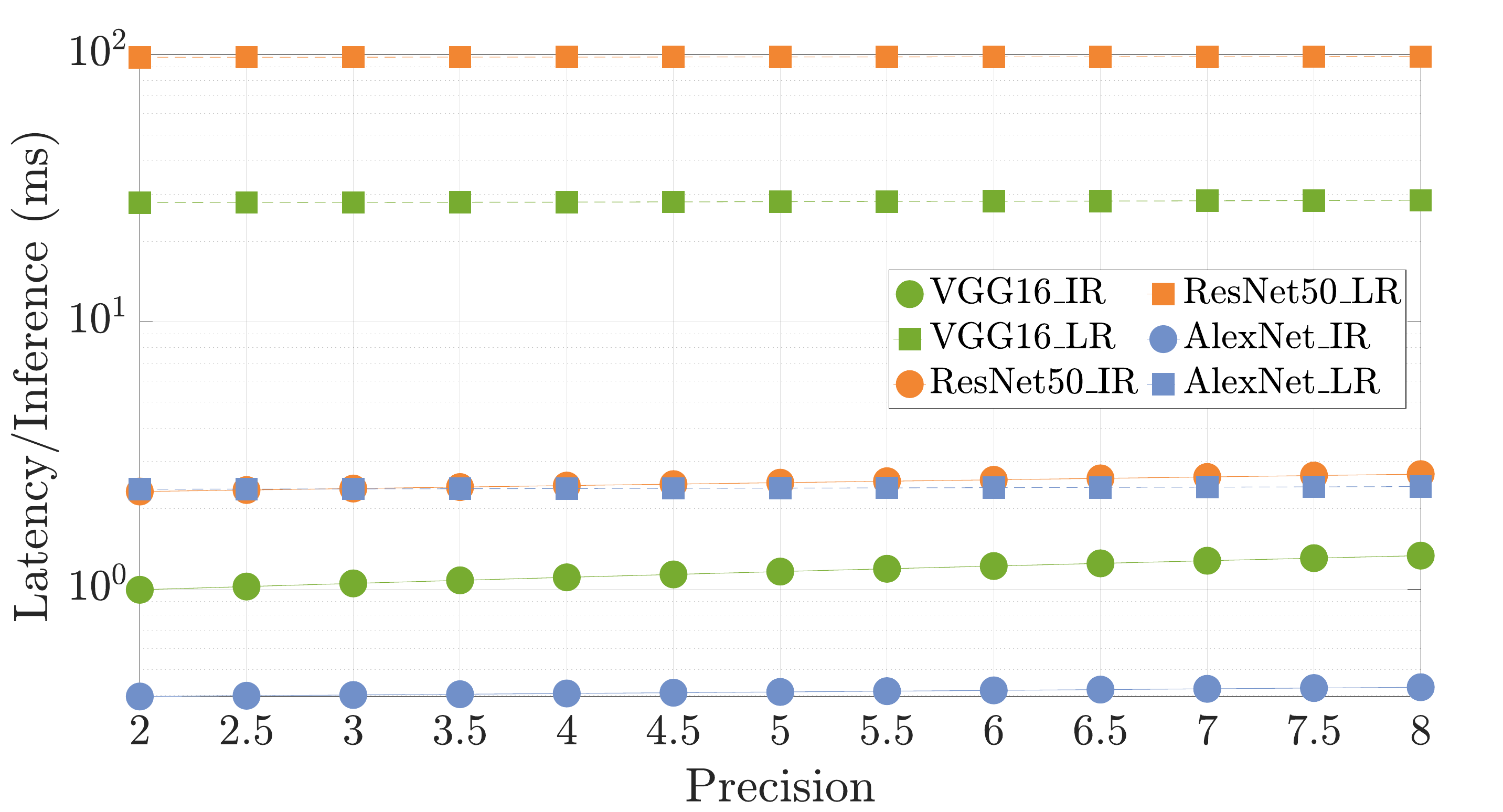}}
\hfil
\subfloat[]
{\includegraphics[width=0.333\textwidth]
{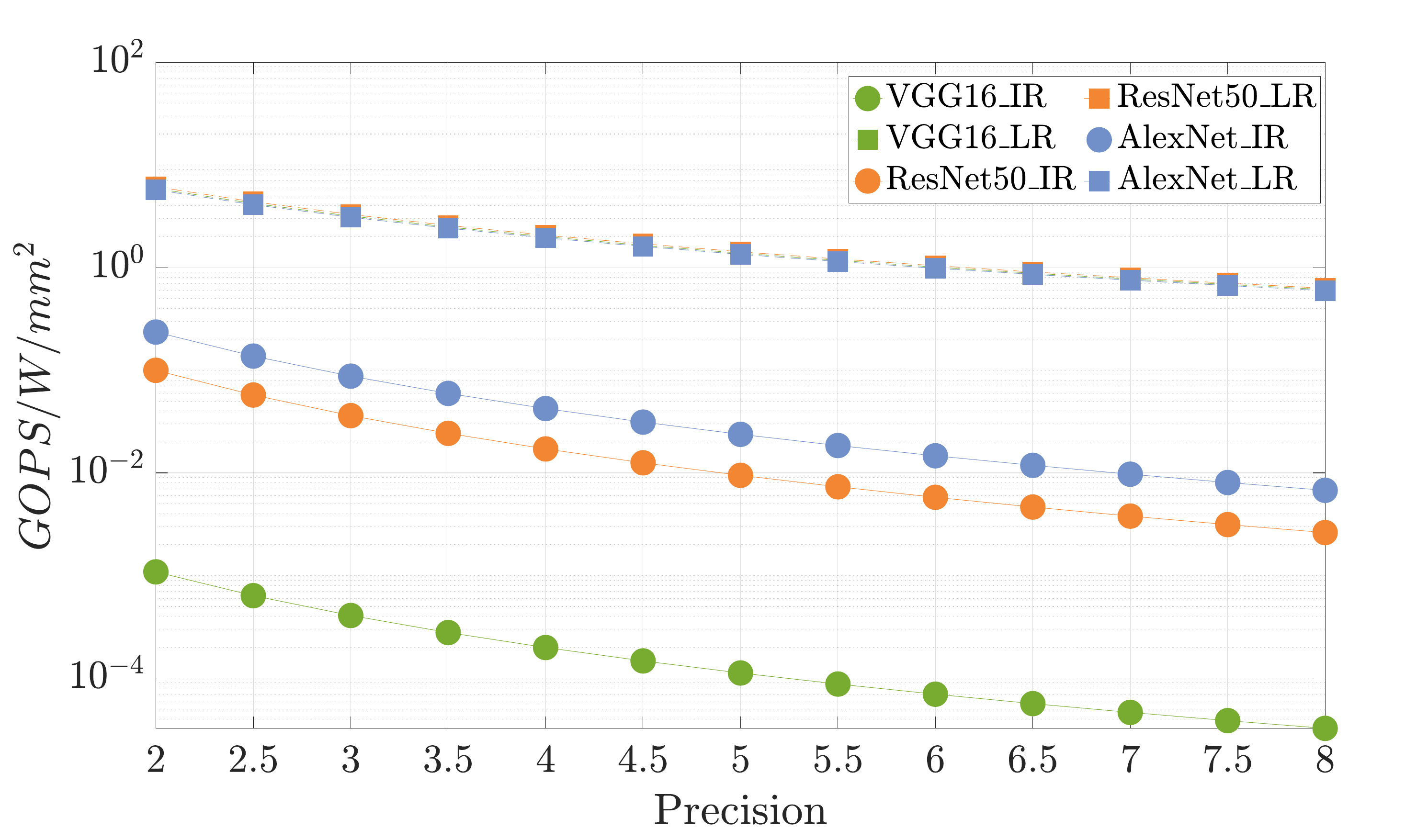}}
\caption{(a) Energy/inference, (b) latency/inference, and (c) $GOPS/w/mm^2$ vs average precision. \label{results2}}
    \vspace{-0.15in}
\end{figure*}


\begin{figure}[ht!]
    \centering
    \subfloat[]{\includegraphics[width=0.4\columnwidth]{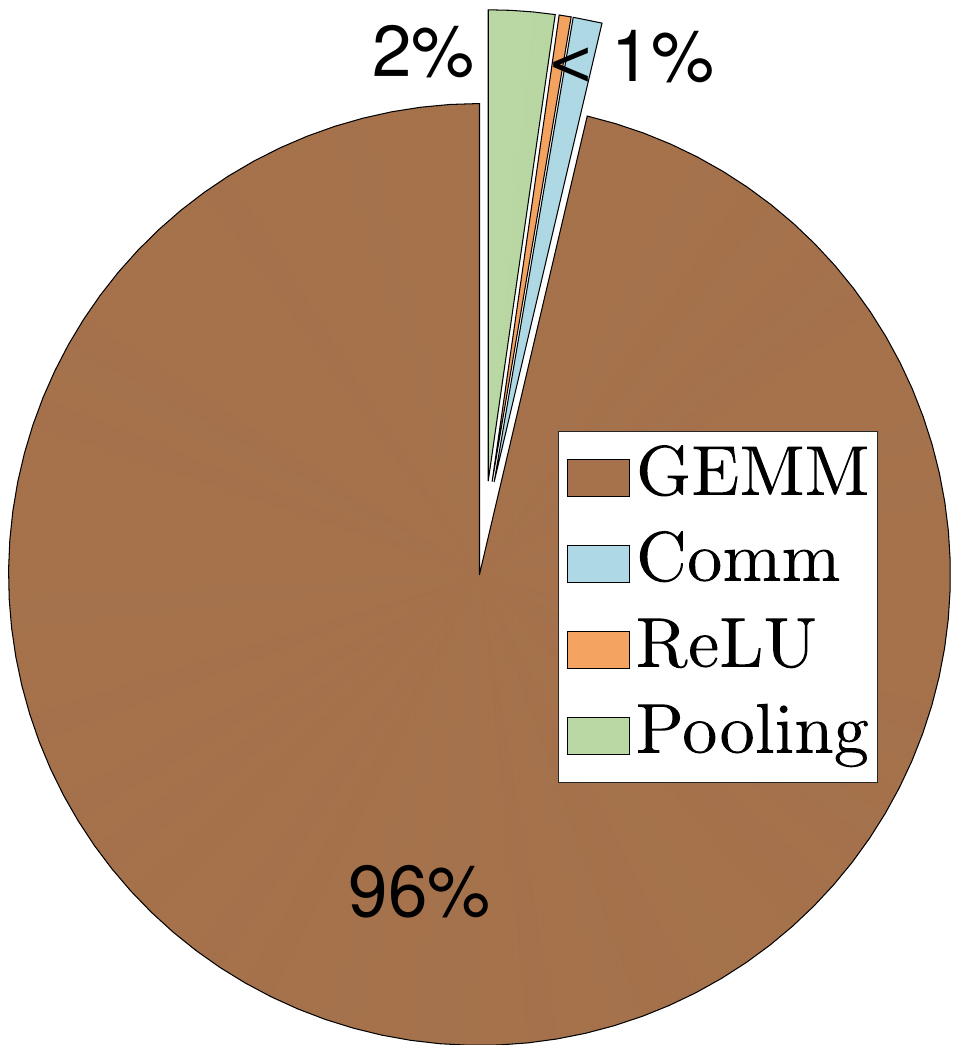}\label{energy_breakdown}}
\hfil
\subfloat[]{\includegraphics[width=0.39\columnwidth]{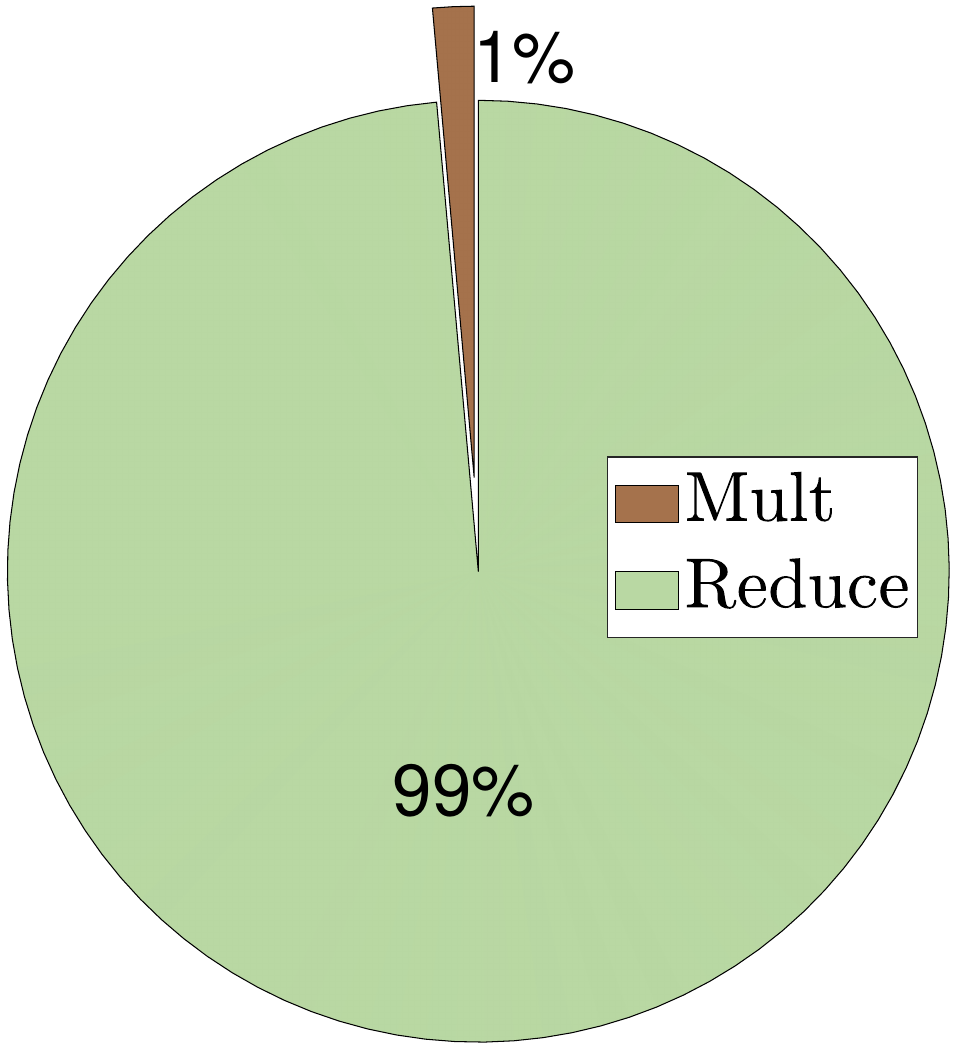}\label{latency_breakdown}}
\hfil
    \caption{Breakdowns of a-) energy and b-) GEMM latency.}
    \vspace{-0.2in}
\end{figure}

\section{Results and Discussion}
\subsection{Design Space Exploration}
In this section, we report the performance metrics of end-to-end inference on BF-IMNA by varying 1-) the technology type: RRAM vs SRAM, 2-) the mixed precision configuration across layers, and 3-) the voltage supplying the cells.

\noindent \textbf{ReRAM Versus SRAM:}
We estimate the performance of end-to-end inference on BF-IMNA using ReRAM-based and SRAM-based CAM cells. Fig. \ref{rram-sram-ratio} shows the energy/latency (ReRAM to SRAM) ratios for different fixed-precisions across all layers of VGG16. We notice that for all precisions, SRAM has lower energy and latency. This is because the SRAM cells require $4$ orders of magnitude less energy to write and require half the cycles to write compared to ReRAM cells. Energy ratios keep decreasing: $80.9\times$, $72.9\times$, $68.9\times$, $66.6\times$, $65.0\times$, $63.9\times$, and $63.1\times$ as precision increases between $2$ and $8$. This phenomenon is because a-) as precision increases, comparison energy becomes the bottleneck and not the write energy (for every pair of columns we do $4$ comparisons and $1.5$ writes on average, and higher precision means more columns to evaluate), and b-) the comparison energy is similar in both technologies. For latency, the ratios remain almost constant $\backsim 1.85\times$ as the precision increases, because the latency bottleneck is the reduction operation (as will be shown below) which relies on the number of rows in the AP rather than the number of columns (affected by precision). These trends are consistent with what we obtain in the mixed-precision study (see below). ReRAM technology is beneficial when area is a concern as it offers $4.4\times$ area savings (to support $8$-bit precision). We also note that AlexNet and ResNet50 exhibit similar ratio trends. In the rest of the experiments, we rely on the SRAM technology for it is more energy and latency-efficient. While we have not tested other resistive AP cells in our simulator, like PCMs \cite{wong2010phase} or FeFETs \cite{muller2012ferroelectricity}, it is very easy to extend our framework to perform a similar analysis for these technologies.

\noindent \textbf{Per-layer Mixed-precision:} Figs. \ref{results2} (a), (b), and (c) show the SRAM-based BF-IMNA energy, latency, and $GOPS/w/mm^2$ for end-to-end inference of ImageNet on VGG16, AlexNet, and ResNet50 with varying average precision. Average precision is calculated across layers, where each layer can have a different precision. Particularly, we evaluate the performance of several mixed-precision per-layer combinations, each of which yields a specific average precision value. The mean performances across the combinations with similar average precision are reported in Fig. \ref{results2}. For each combination, the average precision is calculated using the same method outlined in Table \ref{hawq-v3-res}. We note that BF-IMNA estimates the end-to-end inference performance metrics including on-chip communication. In Fig. \ref{results2} (a) the energy per inference of VGG16 exceeds ResNet50 which in turn exceeds AlexNet. This is because the number of MAC operations of VGG16 ($15.5G$) exceeds ResNet50 ($4.14G$) which exceeds AlexNet ($0.72G$). The slight difference in energy between the IR and LR hardware configurations of each mapped model (up to $0.04J$) is due to slight differences in data movement. As the average precision increases from 2 to 8, we see a non-linear increase in energy, since more cells would be processed. For example, for the LR configuration, the energy/inference on ResNet50 increases from $0.009$ to $0.095J$ ($10.5\times$) as average precision increases from 2 to 8 bits. Figs. \ref{energy_breakdown} and \ref{latency_breakdown} represent the general trends of the total energy and GEMM latency breakdowns for the three models. Fig. \ref{energy_breakdown} reveals that GEMM and pooling are the main energy bottlenecks. This is why the energy trends observed for all models vs average precision are closest to the combination of trends seen in Figs. \ref{runtimes} (b), (c), and (d) vs the precision, $M$.

Clearly from Fig. \ref{results2} (b), changing the average precision does not impact the latency significantly. This phenomenon is explained by Fig. \ref{latency_breakdown}, which shows that the latency bottleneck of GEMM is the reduction and not the multiplication. Reduction is performed between pairs of rows sequentially while multiplication is performed between columns bit-serially. Since precision affects the number of columns only, latency increases slightly due to the contribution of multiplication. The latency trend observed for all models versus average precision is closest to the trend seen in Fig. \ref{runtimes} (a) versus "M". For all models, the LR-based mapping incurs higher latency than the IR-based mapping as the former necessitates a folding in the time domain. Particularly, the latency overhead incurred by switching from the IR-based to the LR-based mapping is up to $42\times$, $28\times $, and $6\times$ for ResNet50, VGG16 and AlexNet respectively. For each hardware configuration, the latency taken by ResNet50 exceeds VGG16 which exceeds AlexNet. For IR-based mapping, we exploit the maximum degree of intra-layer parallelism, and the bottleneck becomes the sequential part of the inference, which is determined by the number of layers. LR-based mapping has the same bottleneck even though we are folding in time.

Fig. \ref{results2} (c) presents the $GOPS/w/mm^2$ for an end-to-end inference. We calculate the effective throughput as follows $GOPS=\#GigaOperations/latency$. The effective energy efficiency is the throughput divided by power, $GOPS/W$, for each model. Effective energy-area efficiency is energy efficiency divided by area, $GOPS/W/mm^2$, for each model. This metric is independent of latency, and the higher the better. For the same model, the LR-based mapping has a higher $GOPS/W/mm^2$ compared to its IR-based mapping. From the energy figure, we have seen that energy does not change significantly between the two hardware configurations for the same average precision and the same model. An LR mapping yields a better area efficiency than LR mapping, thus LR-based mapping has higher energy-area efficiency than IR-based. For a single configuration and model, increasing the average precision increases the area and energy so $GOPS/W/mm^2$ decreases. Also, IR-based configurations have up to 4 orders of magnitude lower energy-area efficiency due to the huge area. The LR results for all models are close so Fig. \ref{results2} (c) shows them as nearly overlapping. For one average precision and LR-based mapping, the maximum variation between two workloads is $7.13\%$.
For LR setting, the area is fixed for all workloads and is chosen such that the utilization is near $100\%$ for all three workloads. $GOPS/W/mm^2$ is affected by the scaling of GOPs and power. These metrics scale similarly between workloads with LR-based mapping, so their ratio remains almost constant as we change the workload (for example changing workloads from AlexNet to ResNet50: on average GOPs scales by $7.16\times$ and power scales by $7.55\times$). The comparison between LR and IR configurations reveals a trade-off between latency and energy-area efficiency: while an infinite resources configuration is faster, it is less energy-area efficient and unrealistic to implement. The compromise in latency for LR configuration does not impact BF-IMNA's overall performance when compared to other SOTA works as will be demonstrated in V.B and V.C. We deduce that overall, lower average precisions in a mixed-precision scenario are more beneficial for BF-IMNA.

\noindent \textbf{Voltage Scaling:} To try to further increase BF-IMNA's energy efficiency, we apply voltage scaling. Previous work showed that in scenarios where "approximate" computing is acceptable, the supply voltage in an SRAM-based AP can be scaled down to $0.5V$, causing the write energy per cell, to decrease from $0.24fJ$ to $0.06fJ$ at an expense of increasing the average probability of error in the cell from $0$ to $0.021$ \cite{yantir2018efficient}. We used our simulator to apply voltage scaling to BF-IMNA. Voltage scaling can impact end-to-end accuracy, but our results revealed insignificant energy savings with voltage scaling (up to $0.06\%$ less energy for all workloads), so it was not worth exploring the effect on end-to-end accuracy. The main reason for the insignificant energy savings with voltage scaling is that now that the write energy is in the sub-fJ, the compare energy's contribution to the total energy is more prominent to the point that reducing the write energy, even more, does not produce any significant benefit, especially with increasing precision. Our reported results elsewhere in all figures and tables assume no voltage scaling.

\begin{table*}[ht!]
\caption{Bit fluid BF-IMNA implements mixed-precision inference on ResNet18 based on HAWQ-V3 \cite{yao2021hawq}.}
\label{hawq-v3-res}
\centering
\resizebox{1\textwidth}{!}{
\begin{tabular}{|c|c|c|c|c|c|c|c|}
\hline
\begin{tabular}[c]{@{}c@{}}\textbf{Latency}\\ \textbf{Constraint}\end{tabular} & \begin{tabular}[c]{@{}c@{}}\textbf{Per Layer Bitwidth}\\ \textbf{(weight and activation)}\end{tabular} & \begin{tabular}[c]{@{}c@{}}\textbf{Average}\\ \textbf{Bitwidth} \end{tabular}& 
\begin{tabular}[c]{@{}c@{}} \textbf{Normalized} \\\textbf{Energy} \end{tabular}& \begin{tabular}[c]{@{}c@{}}\textbf{Normalized}\\ \textbf{Latency} \end{tabular}& \textbf{EDP (J.sec)}                        & \textbf{Size (MB)}                   & \begin{tabular}[c]{@{}c@{}}\textbf{Top 1\%}\\ \textbf{Accuracy}         \end{tabular}     \\ \hline
\textbf{-}                  & \textbf{19x\{4\}}                                   & \textbf{4}                & \textbf{3.29}    & \textbf{1.004}    & \cellcolor[HTML]{A9D18E}\textbf{0.58} & \cellcolor[HTML]{A9D18E}\textbf{5.6} & \textbf{68.45}                         \\ \hline
\textbf{High}               & \{8,8,8,8,8,8,8,8,4,8,8,8,8,4,8,4,8,8,4\}           & 7.16                       & 1.13             & 1.001             & 1.69                                  & 8.7                                  & \cellcolor[HTML]{9EC3E7}70.4           \\ \hline
\textbf{Medium}             & \{8,8,8,8,8,4,8,8,4,8,8,8,4,4,8,4,8,4,4\}           & 6.53                           & 1.22            & 1.002             & 1.56                                   & 7.2                                  & 70.34                                  \\ \hline
\textbf{Low}                & \{8,8,8,4,8,4,8,4,4,4,4,4,4,4,4,4,4,4,4\}           & 5.05                   & 1.90            & 1.004             & \cellcolor[HTML]{9EC3E7}1.00          & \cellcolor[HTML]{9EC3E7}6.1          & 68.56                                  \\ \hline
\textbf{-}                  & \textbf{19x\{8\}}                                   & \textbf{8}                      & \textbf{1}            & \textbf{1}             & \textbf{1.91}                         & \textbf{11.2}                        & \cellcolor[HTML]{A9D18E}\textbf{71.56} \\ \hline
\end{tabular}}

\end{table*}


\begin{table}[hpb!]
\vspace{-0.15in}
    \centering
    \caption{Performance comparison with SOTA frameworks.}
    \label{sota}
    \resizebox{\columnwidth}{!}{
    \begin{tabular}{|c|c|c|c|c|c|c|}
        \hline
        \textbf{Framework} & \textbf{\begin{tabular}[c]{@{}c@{}}Technology \\Node\end{tabular}} & \textbf{\begin{tabular}[c]{@{}c@{}} Frequency \\(GHz)\end{tabular}} & \textbf{\begin{tabular}[c]{@{}c@{}}Precision \\(Bits)\end{tabular}} & 
        \textbf{GOPS} & \textbf{GOPS/W} \\
        \hline
        H100 GPU \cite{h100} & CMOS (TSMC 4N) & 1.83 & 8 
        & 1979000 & 2827 \\
        \hline
        TPUv4 \cite{tpuv4} & CMOS (7nm) & 1.05 & 8 
        & 275000 & 1432 \\
        \hline
        \cite{valavi201964} & CMOS (65nm) & 0.1 & 1 
        & 18876 & 866000 \\
        \hline
        \cite{sim201614} & CMOS (65nm) & 0.125 & 16 
        & 64 & 1422 \\
        \hline
        DaDianNao \cite{chen2014diannao} & CMOS (32nm) & 606 & 16 
        & 5584 & 278 \\
        \hline
        ISAAC \cite{shafiee2016isaac} & \begin{tabular}[c]{@{}c@{}}CMOS (32nm)-\\Memristive\end{tabular} & 1.2 & 16 
        & 40907 & 622 \\
        \hline
        PipeLayer \cite{song2017pipelayer} & \begin{tabular}[c]{@{}c@{}}CMOS (50nm)-\\Memristive\end{tabular} & - & 16 
        & 122706 & 143 \\
        \hline
        IMCA \cite{yantir2021imca}& CMOS (65nm) & 1 & 8 
        & 3 & 4630 \\
        \hline
        PUMA \cite{ankit2019puma} & \begin{tabular}[c]{@{}c@{}}CMOS (32nm)-\\Memristive\end{tabular} & 1 & 16 
        & 52310 & 840 \\
        \hline
        \rowcolor{gray!30}BF-IMNA\_1b & CMOS (16nm) & 1 & 1 
        & 2808686 & 22879 \\
        \hline
        \rowcolor{gray!30}BF-IMNA\_8b & CMOS (16nm) & 1 & 8 
        & 140434 & 641 \\ 
        \hline
        \rowcolor{gray!30}BF-IMNA\_16b & CMOS (16nm) & 1 & 16 
        & 41654 & 170 \\ 
        \hline
    \end{tabular}} 
\end{table}
\subsection{Bit Fluidity and Scalability}
BF-IMNA is capable of mixed-precision inference, a topic of growing interest \cite{rakka2022mixed}. It stands out as the sole in-memory hardware capable of dynamic mixed-precision, providing tangible hardware savings in energy and latency. Owing to the frugality of the bit-serial operation, the AP seamlessly shifts between precisions by computing as many bits as needed, without the hardware overhead of run-time reconfiguration. GPUs, though capable of static mixed precision (and possibly dynamic) support, demand significant programming effort for specialized kernels to achieve energy efficiency \cite{wang2020packing}.

To demonstrate BF-IMNA's bit fluidity, we implement the mixed-precision inference scenario on ResNet18, which was defined by HAWQ-V3 \cite{yao2021hawq}. In particular, HAWQ-V3's optimization algorithm chooses the precision per layer for ResNet18 to be either INT4 or INT8 according to three latency constraints: low, medium, and high. The reasoning behind the INT4/INT8 combination is defined by HAWQ-V3 as follows: fixed-precision INT4 quantization could degrade accuracy, fixed-precision INT8 quantization slows down inference (compared to INT4), so using mixed-precision strikes a balance between accuracy and speed. On one hand, HAWQ-V3's framework provides the accuracy for the mixed-precision configuration meeting a certain constraint, and on the other hand, we take the per-layer weight and activation precisions specified by HAWQ-V3 and estimate the hardware cost of performing the end-to-end inference on BF-IMNA (with LR configuration and parameters defined in Table \ref{2DAP_NNAparam}). Table \ref{hawq-v3-res} shows the normalized energy and latency (with respect to INT8), and EDPs reported by our simulator for the mixed-precision configurations found by HAWQ-V3 for different latency budgets. We also report metrics with fixed precision INT4 (top row) and INT8 (bottom row) implementations. The model size and accuracy are adopted from \cite{yao2021hawq}. The results reveal that fixed-precision INT4 achieves the lowest EDP (best) and smallest model size, as expected, but at an expense of $3.11\%$ degradation in accuracy compared to INT8. With a low latency constraint, BF-IMNA reports an EDP of $1J.sec$, lower (better) than EDPs reported for the other latency constraints and closest to fixed precision INT4's EDP, with a $3\%$ degradation in accuracy (less than INT4). The mixed-precision with a high latency constraint achieves the closest accuracy to INT8 (just $1.16\%$ degradation) and reports an EDP of $1.69J.sec$, $1.13\times$ better than INT8. We see that the mixed-precision implementation strikes the balance between accuracy and efficiency. It is worth noting that while HAWQ-V3 finds the precisions for each latency budget separately, BF-IMNA allows switching between the three mixed-precision configurations dynamically, as imposed by the changing run-time resource requirements. For future work, we hope to integrate BF-IMNA as hardware-in-the-loop for "real" (not simulated, aka "fake" as defined by \ref{hawq-v3-res}) mixed-precision, where our simulator's metrics are used to optimize for the mixed-precision configuration choice.


An AP's versatility allows it to perform any functionality in the end-to-end CNN inference, by simple runtime reprogrammability by the controller without hardware overhead. This versatility renders the AP a modular, configurable architecture that can be easily scaled-out with multiple boards and scaled-up with multiple chips per board to form chiplets \cite{loh2021understanding}. Notably, while Section V presents results for a batch size of 1, BF-IMNA readily enables inter-batch pipelining to achieve higher throughput.

\begin{figure}[b!]
\centering
\vspace{-0.15in}
{\includegraphics[width=0.9\columnwidth]{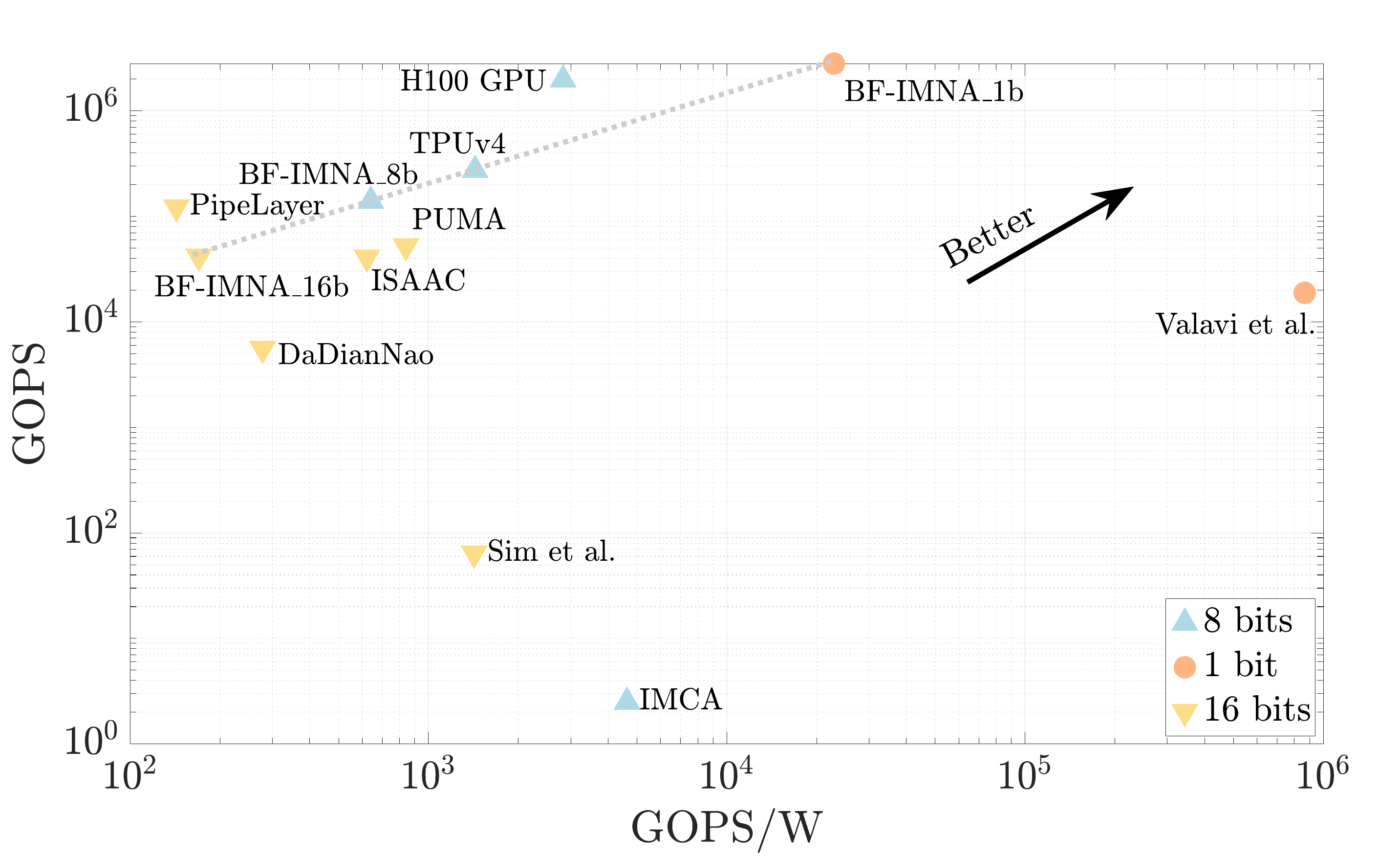}}
\hfil
\caption{GOPS vs GOPS/W. BF-IMNA has high throughput and energy efficiency.}
\label{gops_gops_w}
\end{figure}

\subsection{Comparison with State-of-the-Art}
The results of SOTA works are reported in Table \ref{sota}. We note that for a fair comparison, we assume only convolution is performed when calculating GOPS and energy efficiency, and we report peak values \cite{smagulova2023resistive} of BF-IMNA for fixed-precisions of 1/8/16 configurations. We also consider the buffering needed (from CAPs to MAPs), similar to SOTA works \cite{shafiee2016isaac, song2017pipelayer}. H100 \cite{h100} reports the highest throughput ($1979000GOPS$) for a precision of 8, and with an area of $814mm^2$ and energy efficiency of $2827GOPS/W$. H100 has an energy-area efficiency of $3GOPS/W/mm^2$ For the same precision BF-IMNA achieves $8GOP/W/mm^2$ ($2.7\times$ better than H100), with a throughput of $140434GOPS$. \cite{valavi201964} is the SOTA work with the highest energy efficiency ($86600GOPS/W$), reported for a precision of 1 and a throughput of $18876GOPS$. For a precision of 1, BF-IMNA reports a higher throughput of $2808686$, $149\times$ better than \cite{valavi201964}, with a $38\times$ lower energy efficiency. We also note that this SOTA work is a matrix-multiplication MACRO capable of performing only convolution and not end-to-end inference (unlike BF-IMNA).

PipeLayer \cite{song2017pipelayer}, ISAAC \cite{shafiee2016isaac}, and \cite{ankit2019puma} are the SOTA works which perform end-to-end acceleration of CNNs for a precision of 16. We note that at this precision, and due to the bit-serial word parallel mode of operation of BF-IMNA, our proposed accelerator loses its benefit and reports generally low throughput and energy efficiency. Compared to ISAAC, BF-IMNA has $1.02\times$ higher throughput and $3.66\times$ lower energy efficiency. Compared to PipeLayer, BF-IMNA has $2.95\times$ lower throughput and $1.19\times$ higher energy efficiency. PUMA has $1.26\times$ and $4.95\times$ better throughput and energy efficiency than BF-IMNA respectively. We argue, however, that INT8 is sufficient for CNNs to produce near-SOTA accuracy \cite{jacob2018quantization}. For INT8, BF-IMNA achieves better throughput and energy efficiency than ISAAC and PipeLayer, and better throughput than PUMA with comparable energy efficiency. We note that while BF-IMNA might not be the most energy-efficient accelerator or the accelerator with the highest throughput, it is the only end-to-end inference accelerator that is capable of supporting static and dynamic mixed-precision, and still delivers performance that is comparable to current SOTA accelerators as shown in Fig. \ref{gops_gops_w}. 
\subsection{Limitations}
This work shows initial potential for relying on APs for mixed-precision CNN inference, but we realize that certain limitations cannot be addressed within the scope of this paper. \noindent\textbf{Mapping:} The mapping used in this paper is standard, im2col. Im2col is used mainly as a proof of concept to demonstrate BF-IMNA’s ability to handle GEMM, a crucial operation in many CNN workloads and widely used in accelerators. Future iterations of BF-IMNA can explore other mappings for further optimization in memory/power efficiency \cite{peng2019optimizing,zhang2020efficient}. \textbf{Supported Workloads:} While BF-IMNA currently supports mixed-precision for CNNs, other ML workloads like Large Language Models (LLMs) can be mapped to APs, and this would be interesting, especially with the increase of mixed-precision efforts for LLMs \cite{li2023llm,huang2024slim}. BF-IMNA can perform all the operations required by generative models, including LLMs. However, to efficiently handle these models, some architectural adjustments are necessary. The energy breakdown for CNNs (Fig. 8) revealed that BF-IMNA’s energy bottleneck is the matrix-multiplications. Given the larger scale of matrix-multiplications in LLMs compared to CNNs, and that matrix-multiplications constitute more than $99\%$ of LLM operations \cite{ivanov2021data}, the current architecture may not be efficient for LLMs. We are exploring the integration of more efficient matrix-multiplication engines in BF-IMNA to address this in a future work. \textbf{Scope of Simulations:} CNN accelerators are coupled to a host CPU, so for BF-IMNA, the CPU bears the burden of im2col transformation overhead. All the overheads associated with the host-accelerator coupling are not considered in this work because BF-IMNA is modeled like SOTA CNN accelerators \cite{chen2019eyeriss,shafiee2016isaac,song2017pipelayer}.

\section{Conclusion}
BF-IMNA is a modular and scalable hardware accelerator based on 2D APs designed for end-to-end CNN inference, suitable for future chiplet design. It represents the first demonstration of a full-fledged CNN accelerator capable of performing static and dynamic mixed-precision. Results have revealed that BF-IMNA demonstrates competitive hardware performance metrics compared to SOTA works for the same fixed-precisions while being frugal to support per-layer mixed-precision. Future work involves integrating optimizations and mapping techniques, extending support to more complex ML workloads beyond CNNs, and integrating BF-IMNA as hardware in the loop for guiding mixed-precision optimization.


\begin{thebibliography}{10}
\providecommand{\url}[1]{#1}
\csname url@samestyle\endcsname
\providecommand{\newblock}{\relax}
\providecommand{\bibinfo}[2]{#2}
\providecommand{\BIBentrySTDinterwordspacing}{\spaceskip=0pt\relax}
\providecommand{\BIBentryALTinterwordstretchfactor}{4}
\providecommand{\BIBentryALTinterwordspacing}{\spaceskip=\fontdimen2\font plus
\BIBentryALTinterwordstretchfactor\fontdimen3\font minus \fontdimen4\font\relax}
\providecommand{\BIBforeignlanguage}[2]{{%
\expandafter\ifx\csname l@#1\endcsname\relax
\typeout{** WARNING: IEEEtranS.bst: No hyphenation pattern has been}%
\typeout{** loaded for the language `#1'. Using the pattern for}%
\typeout{** the default language instead.}%
\else
\language=\csname l@#1\endcsname
\fi
#2}}
\providecommand{\BIBdecl}{\relax}
\BIBdecl

\bibitem{ankit2019puma}
A.~Ankit, I.~E. Hajj, S.~R. Chalamalasetti, G.~Ndu, M.~Foltin, R.~S. Williams, P.~Faraboschi, W.-m.~W. Hwu, J.~P. Strachan, K.~Roy \emph{et~al.}, ``Puma: A programmable ultra-efficient memristor-based accelerator for machine learning inference,'' in \emph{Proceedings of the Twenty-Fourth International Conference on Architectural Support for Programming Languages and Operating Systems}, 2019, pp. 715--731.

\bibitem{bulat2021bit}
A.~Bulat and G.~Tzimiropoulos, ``Bit-mixer: Mixed-precision networks with runtime bit-width selection,'' in \emph{Proceedings of the IEEE/CVF International Conference on Computer Vision}, 2021, pp. 5188--5197.

\bibitem{chen2014diannao}
T.~Chen, Z.~Du, N.~Sun, J.~Wang, C.~Wu, Y.~Chen, and O.~Temam, ``Diannao: A small-footprint high-throughput accelerator for ubiquitous machine-learning,'' \emph{ACM SIGARCH Computer Architecture News}, vol.~42, no.~1, pp. 269--284, 2014.

\bibitem{chen2019eyeriss}
Y.-H. Chen, T.-J. Yang, J.~Emer, and V.~Sze, ``Eyeriss v2: A flexible accelerator for emerging deep neural networks on mobile devices,'' \emph{IEEE Journal on Emerging and Selected Topics in Circuits and Systems}, vol.~9, no.~2, pp. 292--308, 2019.

\bibitem{collobert2002torch}
R.~Collobert, S.~Bengio, and J.~Mari{\'e}thoz, ``Torch: a modular machine learning software library,'' Idiap, Tech. Rep., 2002.

\bibitem{dally2020domain}
W.~J. Dally, Y.~Turakhia, and S.~Han, ``Domain-specific hardware accelerators,'' \emph{Communications of the ACM}, vol.~63, no.~7, pp. 48--57, 2020.

\bibitem{dou2020exploiting}
Z.-Y. Dou, X.~Wang, S.~Shi, and Z.~Tu, ``Exploiting deep representations for natural language processing,'' \emph{Neurocomputing}, vol. 386, pp. 1--7, 2020.

\bibitem{foster1976content}
C.~C. Foster, \emph{Content addressable parallel processors}.\hskip 1em plus 0.5em minus 0.4em\relax John Wiley \& Sons, Inc., 1976.

\bibitem{fouda2022memory}
M.~E. Fouda, H.~E. Yant{\i}r, A.~M. Eltawil, and F.~Kurdahi, ``In-memory associative processors: Tutorial, potential, and challenges,'' \emph{IEEE Transactions on Circuits and Systems II: Express Briefs}, vol.~69, no.~6, pp. 2641--2647, 2022.

\bibitem{goel2020survey}
A.~Goel, C.~Tung, Y.-H. Lu, and G.~K. Thiruvathukal, ``A survey of methods for low-power deep learning and computer vision,'' in \emph{2020 IEEE 6th World Forum on Internet of Things (WF-IoT)}.\hskip 1em plus 0.5em minus 0.4em\relax IEEE, 2020, pp. 1--6.

\bibitem{he2016deep}
K.~He, X.~Zhang, S.~Ren, and J.~Sun, ``Deep residual learning for image recognition,'' in \emph{Proceedings of the IEEE conference on computer vision and pattern recognition}, 2016, pp. 770--778.

\bibitem{huang2024slim}
W.~Huang, H.~Qin, Y.~Liu, Y.~Li, X.~Liu, L.~Benini, M.~Magno, and X.~Qi, ``Slim-llm: Salience-driven mixed-precision quantization for large language models,'' \emph{arXiv preprint arXiv:2405.14917}, 2024.

\bibitem{hussain2022design}
H.~Hussain, P.~Tamizharasan, and C.~Rahul, ``Design possibilities and challenges of dnn models: a review on the perspective of end devices,'' \emph{Artificial Intelligence Review}, pp. 1--59, 2022.

\bibitem{ivanov2021data}
A.~Ivanov, N.~Dryden, T.~Ben-Nun, S.~Li, and T.~Hoefler, ``Data movement is all you need: A case study on optimizing transformers,'' \emph{Proceedings of Machine Learning and Systems}, vol.~3, pp. 711--732, 2021.

\bibitem{jacob2018quantization}
B.~Jacob, S.~Kligys, B.~Chen, M.~Zhu, M.~Tang, A.~Howard, H.~Adam, and D.~Kalenichenko, ``Quantization and training of neural networks for efficient integer-arithmetic-only inference,'' in \emph{Proceedings of the IEEE conference on computer vision and pattern recognition}, 2018, pp. 2704--2713.

\bibitem{jia2014caffe}
Y.~Jia, E.~Shelhamer, J.~Donahue, S.~Karayev, J.~Long, R.~Girshick, S.~Guadarrama, and T.~Darrell, ``Caffe: Convolutional architecture for fast feature embedding,'' in \emph{Proceedings of the 22nd ACM international conference on Multimedia}, 2014, pp. 675--678.

\bibitem{tpuv4}
N.~Jouppi, G.~Kurian, S.~Li, P.~Ma, R.~Nagarajan, L.~Nai, N.~Patil, S.~Subramanian, A.~Swing, B.~Towles \emph{et~al.}, ``Tpu v4: An optically reconfigurable supercomputer for machine learning with hardware support for embeddings,'' in \emph{Proceedings of the 50th Annual International Symposium on Computer Architecture}, 2023, pp. 1--14.

\bibitem{judd2016stripes}
P.~Judd, J.~Albericio, T.~Hetherington, T.~M. Aamodt, and A.~Moshovos, ``Stripes: Bit-serial deep neural network computing,'' in \emph{2016 49th Annual IEEE/ACM International Symposium on Microarchitecture (MICRO)}.\hskip 1em plus 0.5em minus 0.4em\relax IEEE, 2016, pp. 1--12.

\bibitem{karam2015emerging}
R.~Karam, R.~Puri, S.~Ghosh, and S.~Bhunia, ``Emerging trends in design and applications of memory-based computing and content-addressable memories,'' \emph{Proceedings of the IEEE}, vol. 103, no.~8, pp. 1311--1330, 2015.

\bibitem{krikelis1994associative}
A.~Krikelis and C.~C. Weems, ``Associative processing and processors,'' \emph{Computer}, vol.~27, no.~11, pp. 12--17, 1994.

\bibitem{krizhevsky2017imagenet}
A.~Krizhevsky, I.~Sutskever, and G.~E. Hinton, ``Imagenet classification with deep convolutional neural networks,'' \emph{Communications of the ACM}, vol.~60, no.~6, pp. 84--90, 2017.

\bibitem{li2018edge}
E.~Li, Z.~Zhou, and X.~Chen, ``Edge intelligence: On-demand deep learning model co-inference with device-edge synergy,'' in \emph{Proceedings of the 2018 Workshop on Mobile Edge Communications}, 2018, pp. 31--36.

\bibitem{li2023llm}
S.~Li, X.~Ning, K.~Hong, T.~Liu, L.~Wang, X.~Li, K.~Zhong, G.~Dai, H.~Yang, and Y.~Wang, ``Llm-mq: Mixed-precision quantization for efficient llm deployment,'' in \emph{The Efficient Natural Language and Speech Processing Workshop with NeurIPS}, vol.~9, 2023.

\bibitem{li2021survey}
Z.~Li, F.~Liu, W.~Yang, S.~Peng, and J.~Zhou, ``A survey of convolutional neural networks: analysis, applications, and prospects,'' \emph{IEEE transactions on neural networks and learning systems}, 2021.

\bibitem{liu2017survey}
W.~Liu, Z.~Wang, X.~Liu, N.~Zeng, Y.~Liu, and F.~E. Alsaadi, ``A survey of deep neural network architectures and their applications,'' \emph{Neurocomputing}, vol. 234, pp. 11--26, 2017.

\bibitem{loh2021understanding}
G.~H. Loh, S.~Naffziger, and K.~Lepak, ``Understanding chiplets today to anticipate future integration opportunities and limits,'' in \emph{2021 Design, Automation \& Test in Europe Conference \& Exhibition (DATE)}.\hskip 1em plus 0.5em minus 0.4em\relax IEEE, 2021, pp. 142--145.

\bibitem{lou2019autoq}
Q.~Lou, F.~Guo, L.~Liu, M.~Kim, and L.~Jiang, ``Autoq: Automated kernel-wise neural network quantization,'' \emph{arXiv preprint arXiv:1902.05690}, 2019.

\bibitem{ma2020memory}
Y.~Ma, Y.~Du, L.~Du, J.~Lin, and Z.~Wang, ``In-memory computing: The next-generation ai computing paradigm,'' in \emph{Proceedings of the 2020 on Great Lakes Symposium on VLSI}, 2020, pp. 265--270.

\bibitem{muller2012ferroelectricity}
J.~M{\"u}ller, E.~Yurchuk, T.~Schl{\"o}sser, J.~Paul, R.~Hoffmann, S.~M{\"u}ller, D.~Martin, S.~Slesazeck, P.~Polakowski, J.~Sundqvist \emph{et~al.}, ``Ferroelectricity in hfo 2 enables nonvolatile data storage in 28 nm hkmg,'' in \emph{2012 symposium on VLSI technology (VLSIT)}.\hskip 1em plus 0.5em minus 0.4em\relax IEEE, 2012, pp. 25--26.

\bibitem{h100}
NVIDIA, ``H100,'' \url{https://resources.nvidia.com}.

\bibitem{peng2019optimizing}
X.~Peng, R.~Liu, and S.~Yu, ``Optimizing weight mapping and data flow for convolutional neural networks on rram based processing-in-memory architecture,'' in \emph{2019 IEEE International Symposium on Circuits and Systems (ISCAS)}.\hskip 1em plus 0.5em minus 0.4em\relax IEEE, 2019, pp. 1--5.

\bibitem{rakka2020design}
M.~Rakka, M.~E. Fouda, R.~Kanj, A.~Eltawil, and F.~J. Kurdahi, ``Design exploration of sensing techniques in 2t-2r resistive ternary cams,'' \emph{IEEE Transactions on Circuits and Systems II: Express Briefs}, vol.~68, no.~2, pp. 762--766, 2020.

\bibitem{rakka2023dt2cam}
M.~Rakka, M.~E. Fouda, R.~Kanj, and F.~Kurdahi, ``Dt2cam: A decision tree to content addressable memory framework,'' \emph{IEEE Transactions on Emerging Topics in Computing}, vol.~11, no.~3, pp. 805--810, 2023.

\bibitem{rakka2022mixed}
M.~Rakka, M.~E. Fouda, P.~Khargonekar, and F.~Kurdahi, ``A review of state-of-the-art mixed-precision neural network frameworks,'' \emph{IEEE Transactions on Pattern Analysis and Machine Intelligence}, 2024.

\bibitem{shafiee2016isaac}
A.~Shafiee, A.~Nag, N.~Muralimanohar, R.~Balasubramonian, J.~P. Strachan, M.~Hu, R.~S. Williams, and V.~Srikumar, ``Isaac: A convolutional neural network accelerator with in-situ analog arithmetic in crossbars,'' \emph{ACM SIGARCH Computer Architecture News}, vol.~44, no.~3, pp. 14--26, 2016.

\bibitem{sharma2018bit}
H.~Sharma, J.~Park, N.~Suda, L.~Lai, B.~Chau, J.~K. Kim, V.~Chandra, and H.~Esmaeilzadeh, ``Bit fusion: Bit-level dynamically composable architecture for accelerating deep neural network,'' in \emph{2018 ACM/IEEE 45th Annual International Symposium on Computer Architecture (ISCA)}.\hskip 1em plus 0.5em minus 0.4em\relax IEEE, 2018, pp. 764--775.

\bibitem{sim201614}
J.~Sim, J.-S. Park, M.~Kim, D.~Bae, Y.~Choi, and L.-S. Kim, ``14.6 a 1.42 tops/w deep convolutional neural network recognition processor for intelligent ioe systems,'' in \emph{2016 IEEE International Solid-State Circuits Conference (ISSCC)}.\hskip 1em plus 0.5em minus 0.4em\relax IEEE, 2016, pp. 264--265.

\bibitem{simonyan2014very}
K.~Simonyan and A.~Zisserman, ``Very deep convolutional networks for large-scale image recognition,'' \emph{arXiv preprint arXiv:1409.1556}, 2014.

\bibitem{sinha2022dnn}
B.~B. Sinha and R.~Dhanalakshmi, ``Dnn-mf: Deep neural network matrix factorization approach for filtering information in multi-criteria recommender systems,'' \emph{Neural Computing and Applications}, vol.~34, no.~13, pp. 10\,807--10\,821, 2022.

\bibitem{smagulova2023resistive}
K.~Smagulova, M.~E. Fouda, F.~Kurdahi, K.~N. Salama, and A.~Eltawil, ``Resistive neural hardware accelerators,'' \emph{Proceedings of the IEEE}, vol. 111, no.~5, pp. 500--527, 2023.

\bibitem{song2017pipelayer}
L.~Song, X.~Qian, H.~Li, and Y.~Chen, ``Pipelayer: A pipelined reram-based accelerator for deep learning,'' in \emph{2017 IEEE international symposium on high performance computer architecture (HPCA)}.\hskip 1em plus 0.5em minus 0.4em\relax IEEE, 2017, pp. 541--552.

\bibitem{theis2017end}
T.~N. Theis and H.-S.~P. Wong, ``The end of moore's law: A new beginning for information technology,'' \emph{Computing in science \& engineering}, vol.~19, no.~2, pp. 41--50, 2017.

\bibitem{valavi201964}
H.~Valavi, P.~J. Ramadge, E.~Nestler, and N.~Verma, ``A 64-tile 2.4-mb in-memory-computing cnn accelerator employing charge-domain compute,'' \emph{IEEE Journal of Solid-State Circuits}, vol.~54, no.~6, pp. 1789--1799, 2019.

\bibitem{van2020bayesian}
M.~van Baalen, C.~Louizos, M.~Nagel, R.~A. Amjad, Y.~Wang, T.~Blankevoort, and M.~Welling, ``Bayesian bits: Unifying quantization and pruning,'' \emph{arXiv preprint arXiv:2005.07093}, 2020.

\bibitem{vasudevan2017parallel}
A.~Vasudevan, A.~Anderson, and D.~Gregg, ``Parallel multi channel convolution using general matrix multiplication,'' in \emph{2017 IEEE 28th international conference on application-specific systems, architectures and processors (ASAP)}.\hskip 1em plus 0.5em minus 0.4em\relax IEEE, 2017, pp. 19--24.

\bibitem{wang2019haq}
K.~Wang, Z.~Liu, Y.~Lin, J.~Lin, and S.~Han, ``Haq: Hardware-aware automated quantization with mixed precision,'' in \emph{Proceedings of the IEEE/CVF conference on computer vision and pattern recognition}, 2019, pp. 8612--8620.

\bibitem{wang2020apq}
T.~Wang, K.~Wang, H.~Cai, J.~Lin, Z.~Liu, H.~Wang, Y.~Lin, and S.~Han, ``Apq: Joint search for network architecture, pruning and quantization policy,'' in \emph{Proceedings of the IEEE/CVF Conference on Computer Vision and Pattern Recognition}, 2020, pp. 2078--2087.

\bibitem{wang2020packing}
X.~Wang and W.~Zhang, ``Packing narrow-width operands to improve energy efficiency of general-purpose gpu computing,'' in \emph{2020 IEEE High Performance Extreme Computing Conference (HPEC)}.\hskip 1em plus 0.5em minus 0.4em\relax IEEE, 2020, pp. 1--7.

\bibitem{wong2010phase}
H.-S.~P. Wong, S.~Raoux, S.~Kim, J.~Liang, J.~P. Reifenberg, B.~Rajendran, M.~Asheghi, and K.~E. Goodson, ``Phase change memory,'' \emph{Proceedings of the IEEE}, vol.~98, no.~12, pp. 2201--2227, 2010.

\bibitem{yantir2018efficient}
H.~E. Yantir, \emph{Efficient acceleration of computation using associative in-memory processing}.\hskip 1em plus 0.5em minus 0.4em\relax University of California, Irvine, 2018.

\bibitem{yantir2018two}
H.~E. Yant{\i}r, A.~M. Eltawil, and F.~J. Kurdahi, ``A two-dimensional associative processor,'' \emph{IEEE Transactions on Very Large Scale Integration (VLSI) Systems}, vol.~26, no.~9, pp. 1659--1670, 2018.

\bibitem{yantir2021imca}
H.~E. Yant{\i}r, A.~M. Eltawil, and K.~N. Salama, ``Imca: An efficient in-memory convolution accelerator,'' \emph{IEEE Transactions on Very Large Scale Integration (VLSI) Systems}, vol.~29, no.~3, pp. 447--460, 2021.

\bibitem{yao2021hawq}
Z.~Yao, Z.~Dong, Z.~Zheng, A.~Gholami, J.~Yu, E.~Tan, L.~Wang, Q.~Huang, Y.~Wang, M.~Mahoney \emph{et~al.}, ``Hawq-v3: Dyadic neural network quantization,'' in \emph{International Conference on Machine Learning}.\hskip 1em plus 0.5em minus 0.4em\relax PMLR, 2021, pp. 11\,875--11\,886.

\bibitem{zha2020hyper}
Y.~Zha and J.~Li, ``Hyper-ap: Enhancing associative processing through a full-stack optimization,'' in \emph{2020 ACM/IEEE 47th Annual International Symposium on Computer Architecture (ISCA)}.\hskip 1em plus 0.5em minus 0.4em\relax IEEE, 2020, pp. 846--859.

\bibitem{zhang2020efficient}
Y.~Zhang, G.~He, G.~Wang, and Y.~Li, ``Efficient and robust rram-based convolutional weight mapping with shifted and duplicated kernel,'' \emph{IEEE Transactions on Computer-Aided Design of Integrated Circuits and Systems}, vol.~40, no.~2, pp. 287--300, 2020.

\end{thebibliography}
\end{document}